# Bilinear flexo-antiferrodistortive coupling in ferroelastics: polar twins, antiphase boundaries and fingerprints of alterelectricity

Eugene A. Eliseev[1], Anna N. Morozovska[2*], Akash Saha[3] and Venkatraman Gopalan[3†]

[1]Frantsevich Institute for Problems in Materials Science, National Academy of Sciences of Ukraine, 3, str. Omeliana Pritsaka, 03142 Kyiv, Ukraine

[2] Institute of Physics of the National Academy of Sciences of Ukraine, 46, Nauki Avenue, 03028 Kyiv, Ukraine

[3] Department of Materials Science and Engineering, Pennsylvania State University, University Park, PA 16802, USA

## Abstract

Using the Landau-Ginsburg-Devonshire approach we show that the linear gradient-type coupling between the electric polarization vector $\vec{P}$ and antiferrodistortive long-range order parameter – pseudovector $\vec{\Phi}$, that has the form of Lifshitz invariant $\frac{1}{2}\left(\vec{P}\cdot\nabla\times\vec{\Phi}-\vec{\Phi}\cdot\nabla\times\vec{P}\right)$ and named "bilinear flexo-antiferrodistortive coupling", can emerge in all antiferrodistortive ferroelastics, since it is symmetry-allowed. Using the four sublattices model, we reveal that the bilinear flexo-antiferrodistortive coupling can induce the sublattice-sensitive polarization at the twin walls and antiphase boundaries in antiferrodistortive ferroelastics without any ferroelectric or antiferroelectric ordering. Since the induced polarization $\overrightarrow{\delta P}$ is perpendicular to $\vec{\Phi}$ and counter-directed in neighboring sublattices with checkerboard-type direction of $\vec{\Phi}$, such structure of $\overrightarrow{\delta P}$ may correspond to the alterelectric-type quadrupolar electric order. However, physical manifestations of the bilinear flexo-antiferrodistortive coupling are invisible in most nanostructured antiferrodistortive ferroelectrics and antiferroelectrics due to the domination of piezoelectric and/or omnipresent linear flexoelectric couplings.

[*] corresponding author, e-mail: anna.n.morozovska@gmail.com, A.N.M. and E.A.E. contributed equally

[†] corresponding author, e-mail: vgopalan@psu.edu, vxg8@psu.edu

## I. Introduction

The flexoelectric-type [1] coupling between the gradients of electric polarization, described by a true polar vector $\vec{P}$, and the antiphase rotation of structural groups, described by an axial pseudovector $\vec{\Phi}$ [2], can lead to the appearance of interfacial polarization at the twin walls, antiphase boundaries and surfaces of antiferrodistortive (AFD) ferroelastics [3, 4]. The known functional form of the flexo-AFD coupling energy is $G_{FAQ} = \frac{\xi_{ijkl}}{2}\left(P_i\Phi_j\frac{\partial\Phi_k}{\partial x_l} - \Phi_i\Phi_j\frac{\partial P_k}{\partial x_l}\right)$, where $\hat{\xi}$ is the fourth-rank true tensor, subscripts $i, j$, $k$ and $l$ are Cartesian indexes varying from 1 to 3 [3, 4]. The energy $G_{FAQ}$, being a true scalar value, is a linear function of $\vec{P}$ and a quadratic function of $\vec{\Phi}$. The linear-quadratic flexo-AFD coupling may contribute significantly to observable interfacial polarization induced by oxygen octahedral rotations at the antiphase boundaries and/or twin walls in $YMnO_3$ [5], $Ca_3Mn_2O_7$ [6], $CaTiO_3$ [7, 8, 9] and $SrTiO_3$ [3, 4, 10], as well as lead to the appearance of versatile spatially modulated structures in multiferroics [11, 12].

The quadratic dependence of $G_{FAQ}$ on $\vec{\Phi}$ appeared principally important for the emergence of interfacial polarization, considered in Refs. [3, 4, 7, 11], because the odd powers of antiphase (i.e., sign-alternating in neighboring sublattices) pseudovector $\vec{\Phi}$ cannot induce the continuous polarization of structural domain walls, surfaces or interfaces in AFD ferroelastics. Therefore, it seemed that it made no sense to consider odd powers of $\vec{\Phi}$ when constructing the sublattice-insensitive flexo-AFD coupling energy.

However, recently Visser et al. [13] introduce "alterelectrics", whose elementary dipole structure is an electric analog of the spin structure in altermagnets [14, 15] or noncollinear antiferromagnets [16] with a spin structure coupled to the inherent structural order (see Fig. **1(a)**). Alterelectricity is based on sign-alternating "sublattice-selective" electric polarization, which displays quadrupolar piezoelectricity (see Fig. **1(b)**). They also established that a counterpart of the spin-separated bands is formed by surface modes, which allow for surface-dependent anisotropic electronic transport analogous to the spintronic applications proposed for altermagnets. Then, the Moire control of alterelectric quadrupolar order has been demonstrated [17]. Fang et al. proposed [18] the concept of alterelectricity, an electrical analogue of altermagnetism, in which two switchable states possess alternating band structures. Using an anisotropic Lieb-lattice model, they derived a general symmetry framework for identifying alterelectricity and predicted realizations of alterelectricity in bilayers with interlayer sliding (such as tetragonal $Ag_2N$ and hexagonal $FeHfI_6$) and Ti-adsorbed $SnP_2S_6$. Importantly, an alterelectric state could arise when a switchable sublattice-selective structural change connects two configurations related by a non-inversion symmetry.

Much earlier, mysterious bright domain walls observed in layered van der Waals antiferrielectric $CuInP_2Se_6$ by the piezoresponse force microscopy (PFM) [19] and were explained theoretically using the model of four sublattices, each of which carry an elementary electric dipole [20, 21]. Possible directions of polarization vector in four sublattices include the sign-alternating checkerboard-type arrangements of dipoles, which polarization $\vec{P}^{(i)}$ can be coupled with the direction the antiphase AFD ordering $\vec{\Phi}^{(i)}$, where the superscript $(i)$ is the sublattice index varying from 1 to 4. The obtained structure, shown **Fig. 1(c),** corresponds to the sublattice-selective "alterelectric" quadrupolar electric order $\vec{P}^{(i)}$ that is related with the direction of AFD order $\vec{\Phi}^{(i)}$ by a non-inversion symmetry operator – clockwise or counterclockwise rotation by an angle $\pm\delta\Phi$. Note that four is the minimal amount of sublattices, which allows to construct 2D checkerboard lattice. Their number can be higher in the 3D case.

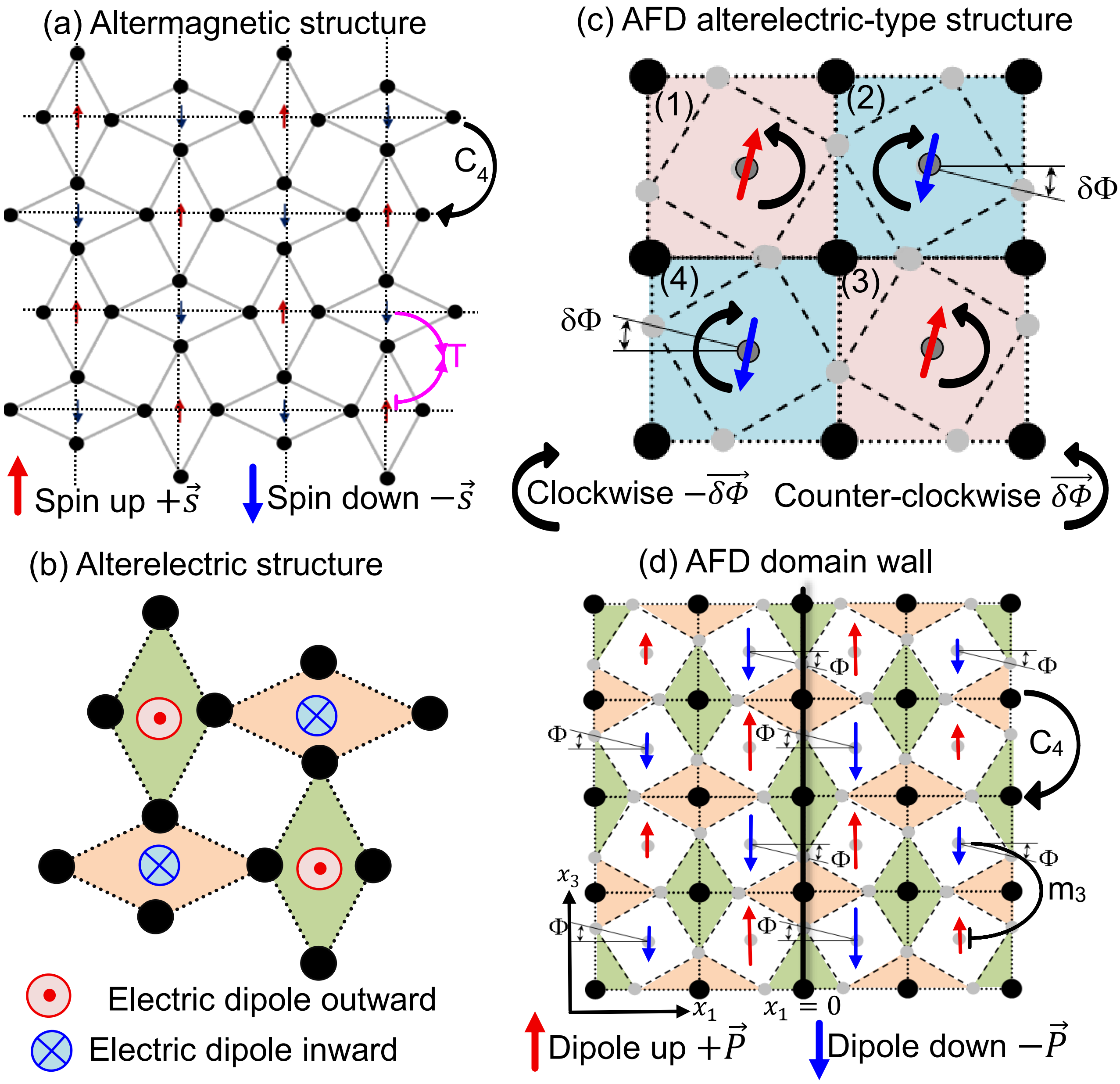

**FIGURE 1. (a)** Sketch of altermagnetic structure, where two neighboring rhombs are connected by the 90-

degree rotation operation $C_4$; opposite spins, shown by red and blue arrows, are connected by time-inversion operation $T$. **(b)** Sketch of alterelectric structure, where electric dipoles in neighboring cells, depicted by red points (inward) and blue crosses (outward) are directed out of the figure plane. Adapted from Visser et al. [13]. **(c)** Sketch of the AFD alterelectric-type structure in four neighboring sublattices, numbered as $i = 1, 2, 3, 4$. Light-grey circles are oxygen atoms; dark-grey circles are B-cations and black circles are A-cations projections on the figure plane of the hypothetic $ABO_3$ structure. The oxygen octahedra rotation of the $i$-th sublattice is described by an axial pseudovector – tilt $\vec{\Phi}^{(i)}$, which clockwise or counterclockwise direction is shown by curved black arrows. The electric polarization of the $i$-th sublattice is described by the polar vector $\vec{P}^{(i)}$, whose in-plane direction "up" or "down" is shown by red and blue arrows. Oxygen atoms in two neighboring sublattices, shown by light-blue and pink colors, are connected by rotation at angle $\pm 2\delta\Phi$. **(d)** Alterelectric-type structure near the AFD domain wall plane $x_1 = 0$. The sublattice electric dipole $\vec{P}^{(i)}$, induced by the bilinear gradient-type coupling, has the form of vector product, $\vec{P}^{(i)} \sim \nabla \times \vec{\Phi}^{(i)}$. Light-olive and light-red neighboring rhombs, which contain oxygen atoms in the corners and A-cation in the center, are connected by the 90-degree rotation operation $C_4$. Oppositely directed electric dipoles, shown by red and blue arrows, are connected by mirror-reflection operation $m_3$: $x_3 \rightarrow -x_3$.

Using the four sublattices model and Landau-Ginsburg-Devonshire (LGD) approach, in this work we demonstrate that the linear gradient-type coupling between the electric polarization vector $\vec{P}^{(i)}$ and AFD order – axial vector $\vec{\Phi}^{(i)}$, whose Lifshitz invariant has the form of vector product $\frac{\lambda^{(ij)}}{2}\left(\vec{P}^{(i)} \cdot \nabla \times \vec{\Phi}^{(j)} - \vec{\Phi}^{(j)} \cdot \nabla \times \vec{P}^{(i)}\right)$ and thus can be classified as bilinear flexo-AFD coupling, can emerge in all AFD ferroelastics since it is symmetry-allowed. We reveal that the bilinear flexo-AFD coupling induces the sublattice-selective counter-directed polarization at the twin walls and antiphase boundaries in antiferrodistortive ferroelastics without any ferroelectric or antiferroelectric long-range ordering (see **Fig. 1(d)**). The functional form of the proposed Lifshitz invariant is reminiscent of the Dzyaloshinskii-Moriya invariant in ferri- and antiferromagnets, which flexo-electric analogue leads to the chiral structurers in ferro- and ferrielectric thin films and nanoparticles [22, 23, 24].

Since the induced polarization $\overrightarrow{\delta P}$ is perpendicular to $\vec{\Phi}$, counter-directed in neighboring sublattices with checkerboard-type direction of $\vec{\Phi}$, the structure of $\overrightarrow{\delta P}$, shown in **Fig. 1(d)**, may correspond to the alterelectric-type electric order. Indeed, the sublattice electric dipole $\vec{P}^{(i)}$, induced by the bilinear gradient-type coupling near the AFD domain wall plane $x_1 = 0$, is proportional to the curl of the tilt vector, namely $\overrightarrow{\delta P}^{(i)} \sim \nabla \times \vec{\Phi}^{(i)}$. Light-olive and light-red neighboring rhombs, which contain oxygen atoms in the corners and A-cation projection in the center of a hypothetic $ABO_3$ structure, are connected by the 90-degree rotation operation $C_4$. Oppositely directed electric dipoles, shown by red and blue arrows, are connected by mirror-reflection operation $m_3$: $x_3 \rightarrow -x_3$. Thus, the alterelectric-type structure near the AFD domain wall, shown in **Fig. 1(d)**, is analogous to the well-known altermagnetic structure shown in **Fig. 1(b)**. However, electric dipoles are located at equidistant positions outside the rhombs in the considered hypothetic $ABO_3$ structure. Note that physical manifestations of the bilinear flexo-AFD coupling may remain invisible in most nanostructured AFD ferroelectrics and antiferroelectrics due to the domination of piezoelectric, linear flexoelectric [25] and biquadrartic polar-AFD [26, 27] couplings.

## II. Basic equations

We use the LGD approach to describe the polarization emerging at the 90-degree twin walls (TW) and 180-degree antiphase boundaries (APB) in AFD ferroelastics. An AFD order parameter $\vec{\Phi}^{(i)}$, related to the rotation of atomic groups in a sublattice (e.g., oxygen octahedron tilt angle or rotation angle of chalcogen pyramids) is an axial pseudovector. An emerging electric polarization (EP) $\vec{P}^{(i)}$, related to the dipole moment in a sublattice, is a true polar vector. Both these parameters change at the TWs and APBs.

The bulk part of the LGD free energy $G$ consists of the following contributions:

$$G = G_{EP} + G_{AFD} + G_{HLS} + G_{EL} + G_{ST} + G_{GR} + G_{FL} + G_{FAQ} + G_{FAL} \, . \qquad (1)$$

Here the AFD and EP energies are Landau-Ginzburg series over even powers of the components of $\vec{\Phi}^{(i)}$ and $\vec{P}^{(i)}$:

$$G_{AFD} = b_{kl}^{(ij)} \Phi_k^{(i)} \Phi_l^{(j)} + b_{klqr}^{(ijmn)} \Phi_k^{(i)} \Phi_l^{(j)} \Phi_q^{(m)} \Phi_r^{(n)} + \cdots, \qquad (2a)$$

$$G_{EP} = a_{kl}^{(ij)} P_k^{(i)} P_l^{(j)} + a_{klqr}^{(ijmn)} P_k^{(i)} P_l^{(j)} P_q^{(m)} P_r^{(n)} + \cdots. \qquad (2b)$$

Hereinafter the Einstein summation rule is performed over repeated spatial Cartesian subscripts $\{k, l, q, r\} = \{1,2,3\}$ and over "sublattice" superscripts, i.e., $\{i, j, m, n\} = \{1, 2, 3, 4\}$. Since the conditions of the sublattices equivalence should be fulfilled, and thus the renumbering of the sublattices does not change the free energy.

The biquadrartic polar-AFD coupling energy, also known as Houchmanzadeh-Lajzerowicz-Salje (HLS) coupling energy [26, 28], has the following form

$$G_{HLS} = \eta_{klqr}^{(ijmn)} \Phi_k^{(i)} \Phi_l^{(j)} P_q^{(m)} P_r^{(n)}, \qquad (2c)$$

Elastic, electrostriction and rotational AFD-striction energies are

$$G_{\mathrm{EL}} = \frac{1}{2} c_{ijkl} u_{ij} u_{kl}, \qquad (2d)$$

$$G_{ST} = -\tilde{Q}_{mnkl}^{(ij)} u_{mn} P_k^{(i)} P_l^{(j)} - \tilde{R}_{mnkl}^{(ij)} u_{mn} \Phi_k^{(i)} \Phi_l^{(j)}, \qquad (2e)$$

In Eq.(2e) $u_{ij}$ are the components of elastic strain tensor, $\tilde{Q}_{mnkl}^{(ij)}$ and $\tilde{R}_{mnkl}^{(ij)}$ are the components of electrostriction and AFD-striction [3] tensors, respectively.

The gradient energy contribution, also known as Ginzburg term, to the LGD free energy is:

$$G_{GR} = v_{klqr}^{(ij)} \frac{\partial \Phi_k^{(i)}}{\partial x_l} \frac{\partial \Phi_q^{(j)}}{\partial x_r} + g_{klqr}^{(ij)} \frac{\partial P_k^{(i)}}{\partial x_l} \frac{\partial P_q^{(j)}}{\partial x_r}, \tag{2f}$$

where $v_{klqr}^{(ij)}$ and $g_{klqr}^{(ij)}$ are the components of the AFD order and polarization gradient tensors, respectively.

All above expressions are biquadratic by their nature. Therefore, they are insensitive to the signs of AFD rotation and/or polarization. The evident expressions of $G_{AFD}$, $G_{EP}$, $G_{ST}$, $G_{GR}$ and $G_{FL}$ for the AFD ferroelastics, which have a cubic-symmetry parent phase, are listed in **Supplement S1** [29].

Flexoelectric, linear-quadratic and bilinear flexo-AFD coupling energies, which are sensitive to the sublattice polarization, have the form:

$$G_{FL} = \frac{f_{mjkl}^{(i)}}{2} \left( u_{mj} \frac{\partial P_k^{(i)}}{\partial x_l} - P_k^{(i)} \frac{\partial u_{mj}}{\partial x_l} \right), \tag{3a}$$

$$G_{FAQ} = \frac{\xi_{lmnq}^{(ijk)}}{2} \left( P_l^{(i)} \Phi_m^{(j)} \frac{\partial \Phi_n^{(k)}}{\partial x_q} - \Phi_m^{(j)} \Phi_n^{(k)} \frac{\partial P_l^{(i)}}{\partial x_q} \right), \tag{3b}$$

$$G_{FAL} = \lambda^{(ij)} \frac{\varepsilon_{qkl}}{2} \left( P_q^{(i)} \frac{\partial \Phi_k^{(j)}}{\partial x_l} - \Phi_k^{(j)} \frac{\partial P_q^{(i)}}{\partial x_l} \right). \tag{3c}$$

Hereinafter the superscripts $\{i, j, k\}$ refers to one of four sublattices; $f_{mjkl}^{(i)}$ and $\xi_{lmnq}^{(ijk)}$ are the components of the omnipresent flexoelectric and linear-quadratic flexo-AFD tensors, respectively. They are true tensors, which do not change the sign under the spatial inversion or mirror reflections. The third contribution is a bilinear flexo-AFD coupling energy, $G_{FAL}$, that was not considered in Refs. [3, 4, 7, 11, 20, 21], since it cannot induce any interfacial polarization in the continuum medium limit. The role of $G_{FAL}$ can be important in alterelectric-type materials without any ferroelectric or antiferroelectric long-range order, but with local sublattice-selective checkerboard-type AFD order $\Phi_k^{(j)}$ in the sublattices. The values $\lambda^{(ij)}$ are scalar constants, which are sensitive to the sublattices interaction type (i.e., either nearest-neighbors or next-to-nearest-neighbors), and $\varepsilon_{jkl}$ is an antisymmetric Levi-Civita object that is an axial pseudotensor.

For ferroelastics with the checkerboard-type AFD order in the four sublattices, such as shown in **Fig. 1(c)**, the contribution of the $G_{FL}$ to the macroscopic polarization is absent. The contribution of $G_{FAQ}$ to the sublattice-insensitive (or continuous) polarization, emerging at the TWs and APBs, can be small, since it is proportional to the second power of $\vec{\Phi}$.

The sign-alternating sublattice-sensitive polarizations $P_j^{(i)}$, induced by the bilinear flexo-AFD coupling with the antiphase sublattice-sensitive tilts $\Phi_j^{(i)}$, can be estimated from the minimization of the bilinear flexo-AFD Lifshitz-type invariant $G_{FAL}$,

$$G_{FAL} = \lambda^{(ii)} \frac{\varepsilon_{jkl}}{2}\left(P_j^{(i)} \frac{\partial \Phi_k^{(i)}}{\partial x_l} - \Phi_k^{(i)} \frac{\partial P_j^{(i)}}{\partial x_l}\right). \quad (4)$$

Expression (4) is a simplified form of Eq.(3a), that is allowed by any symmetry of the AFD material. Indeed, the identical form of Eq.(4) is $G_{FAL} = \frac{\lambda^{(ii)}}{2}\left(\vec{P}^{(i)} \cdot \nabla \times \vec{\Phi}^{(i)} - \vec{\Phi}^{(i)} \cdot \nabla \times \vec{P}^{(i)}\right)$, where $\nabla$ is the nabla-operator. Since $\nabla \times \vec{P}^{(i)}$ is a pseudovector and $\nabla \times \vec{\Phi}^{(i)}$ is a true vector, the scalar products $\vec{P}^{(i)} \cdot \nabla \times \vec{\Phi}^{(i)}$ and $\vec{\Phi}^{(i)} \cdot \nabla \times \vec{P}^{(i)}$ are true scalars, which consideration are consistent with arbitrary symmetry of the parent phase.

From Eqs.(3)-(4) one obtains the polarization variation near the TWs and APBs:

$$\delta P_j^{(i)}(\vec{x}) \cong -\zeta_{jmnq}^{(ilk)} \Phi_m^{(l)} \frac{\partial \Phi_n^{(k)}}{\partial x_q} - \chi \varepsilon_{jkl} \frac{\partial \Phi_k^{(i)}}{\partial x_l}. \quad (5)$$

Here we introduced the coupling tensors $\zeta_{jmnq}^{(ilk)} = \left(a_{jr}^{(il)}\right)^{-1} \xi_{rmnq}^{(ilk)}$ and $\chi = \lambda^{(ii)}\left(a_{11}^{(ii)}\right)^{-1}$ for cubic parent phase symmetry, that is assumed for plots calculations in next section.

Assuming that the AFD order parameter is dimensionless (e.g., oxygen octahedra tilt measured in degrees or radians), the bilinear flexo-AFD coupling strength $\chi$ has the dimension of coulomb per meter. The magnitude of $\chi$ can be estimated as $|\chi| \sim \frac{Z_B^*}{d_{AFD}} \sim (0.8 - 40)$ nC/m, where $Z_B^* \sim (0.1 - 1)e_0$ is the effective Born (or Bader) charge, $e_0 = 1.6 \cdot 10^{-19}$C is the elementary charge and $d_{AFD} \sim (4 - 20)$ pm is the effective charge displacement related to the AFD rotation of the corresponding atomic group.

The continuously changing (i.e., macroscopic) polar and AFD-order parameters are related to individual sublattices as $P_j = \frac{1}{4}\sum_{i=1}^{4} P_j^{(i)}$ and $\Phi_j = \frac{1}{4}\sum_{i=1}^{4} \Phi_j^{(i)}$, where the sublattice superscript $i = 1, 2, 3, 4$ and coordinate subscript $j = 1,2,3$. Using the Dzialoshinsky transformation [20], antipolar ordering can be constructed as e.g., $A_j = \frac{1}{4}\left(P_j^{(1)} - P_j^{(2)} + P_j^{(3)} - P_j^{(4)}\right)$. Similarly, the macroscopic rotation $\Phi_j = \frac{1}{4}\sum_{i=1}^{4} \Phi_j^{(i)}$ should be zero, but the combination $\Psi_j = \frac{1}{4}\left(\Phi_j^{(1)} - \Phi_j^{(2)} + \Phi_j^{(3)} - \Phi_j^{(4)}\right)$ can be nonzero due to the sign-alternating AFD-ordering in the neighboring sublattices.

### III. Polarization distribution near TWs and APBs in AFD ferroelastics

Let us consider the checkerboard-type sign-alternating AFD order in the sublattices, namely $\Phi_j^{(1)} = -\Phi_j^{(2)} = \Phi_j^{(3)} = -\Phi_j^{(4)}$ (see **Fig. 1(b)**), whose envelope changes continuously at 90-degree TWs and 180-degree APBs. Corresponding approximate expressions for the $\Phi_j^{(i)}(\vec{x})$ near the TWs and APBs are listed in the first three lines of **Tables I** and **II**, respectively. Details of their derivation are given in **Supplement S1** [29] (see also Refs. [3, 28]). Note that the intrinsic width of easy TWs is much smaller than that of hard TWs, namely $w_{et} \ll w_{ht}$, as well as the intrinsic width of easy APBs is much smaller than that of hard APBs, namely $w_{ea} \ll w_{ha}$ [3]. Approximate expressions for $\delta P_j^{(i)}(\vec{x})$ near the TWs and APBs are listed in the last three lines of **Tables I** and **II**, respectively**.** Details of their derivation are also given in **Supplement S1** [29]**.**

**Table I.** Order parameters near "easy" (i.e., lower energy) and "hard" (i.e., higher energy) 90-degree TWs in the orthorhombic phase of AFD ferroelastics with a cubic parent phase*

| Component | Easy 90-degree TWs | Hard 90-degree TWs |
|---|---|---|
| $\tilde{\Phi}_1^{(i)}$ * | $\Phi_1^{(1)} = \tilde{\Phi}_s \left\{1 + \frac{\mu}{\cosh^2(\tilde{x}_1/w_{et})}\right\}$, <br> $\Phi_1^{(2)} = -\Phi_1^{(1)}$, $\Phi_1^{(3)} = \Phi_1^{(1)}$, $\Phi_1^{(4)} = -\Phi_1^{(1)}$ | $\Phi_1^{(1)} = \tilde{\Phi}_s \tanh\left(\frac{\tilde{x}_1}{w_{ht}}\right)$, <br> $\Phi_1^{(2)} = -\Phi_1^{(1)}$, $\Phi_1^{(3)} = \Phi_1^{(1)}$, $\Phi_1^{(4)} = -\Phi_1^{(1)}$ |
| $\tilde{\Phi}_2^{(i)}$ | $\Phi_2^{(1)} = \tilde{\Phi}_s \tanh\left(\frac{\tilde{x}_1}{w_{et}}\right)$, <br> $\Phi_2^{(2)} = -\Phi_2^{(1)}$, $\Phi_2^{(3)} = \Phi_2^{(1)}$, $\Phi_2^{(4)} = -\Phi_2^{(1)}$ | $\Phi_2^{(1)} = \tilde{\Phi}_s \left\{1 + \frac{\mu}{\cosh^2(\tilde{x}_1/w_{ht})}\right\}$ <br> $\Phi_2^{(2)} = -\Phi_2^{(1)}$, $\Phi_2^{(3)} = \Phi_2^{(1)}$, $\Phi_2^{(4)} = -\Phi_2^{(1)}$ |
| $\Phi_3^{(i)}$ | 0 | 0 |
| $\delta\tilde{P}_1^{(i)}$**, *** | $\eta\tilde{\Phi}_s\tilde{P}_{01}\,\frac{\tanh\left(\frac{\tilde{x}_1}{w_{et}}\right)\left[\tilde{\rho} - 2\mu\left\{1+\frac{\mu}{\cosh^2(\tilde{x}_1/w_{et})}\right\}\right]}{\cosh^2(\tilde{x}_1/w_{et})}$, <br> $\tilde{P}_{01} = -\frac{\tilde{\xi}_{1111}^{(111)}}{w_{et}a_{11}^{(11)}}\tilde{\Phi}_s$, $\tilde{\rho} = \frac{\tilde{\xi}_{1221}^{(111)}}{\tilde{\xi}_{1111}^{(111)}}$, $\eta = \frac{\varepsilon_0\varepsilon_b a_{11}^{(11)}}{1+\varepsilon_0\varepsilon_b a_{11}^{(11)}}$ | $\eta\tilde{\Phi}_s\tilde{P}_{01}\,\frac{\tanh\left(\frac{\tilde{x}_1}{w_{ht}}\right)\left[1 - 2\tilde{\rho}\mu\left\{1+\frac{\mu}{\cosh^2(\tilde{x}_1/w_{ht})}\right\}\right]}{\cosh^2(\tilde{x}_1/w_{ht})}$, <br> $\tilde{P}_{01} = -\frac{\tilde{\xi}_{1111}^{(111)}}{w_{ht}a_{11}^{(11)}}\tilde{\Phi}_s$, $\tilde{\rho} = \frac{\tilde{\xi}_{1221}^{(111)}}{\tilde{\xi}_{1111}^{(111)}}$, $\eta = \frac{\varepsilon_0\varepsilon_b a_{11}^{(11)}}{1+\varepsilon_0\varepsilon_b a_{11}^{(11)}}$ |
| $\delta\tilde{P}_2^{(i)}$ | $\tilde{\Phi}_s\tilde{P}_{02}\,\frac{1+\frac{\mu}{\cosh^2(\tilde{x}_1/w_{et})} - 2\mu\tilde{\epsilon}\tanh^2\left(\frac{\tilde{x}_1}{w_{et}}\right)}{\cosh^2(\tilde{x}_1/w_{et})}$, <br> $\tilde{P}_{02} = -\frac{\tilde{\xi}_{2121}^{(111)}}{w_{et}a_{22}^{(11)}}\tilde{\Phi}_s$, $\tilde{\epsilon} = \frac{\tilde{\xi}_{2211}^{(111)}}{\tilde{\xi}_{2121}^{(111)}}$, | $\tilde{\Phi}_s\tilde{P}_{02}\,\frac{\tilde{\epsilon}\left[1+\frac{\mu}{\cosh^2(\tilde{x}_1/w_{ht})}\right] - 2\mu\tanh^2\left(\frac{\tilde{x}_1}{w_{ht}}\right)}{\cosh^2(\tilde{x}_1/w_{ht})}$, <br> $\tilde{P}_{02} = -\frac{\tilde{\xi}_{2121}^{(111)}}{w_{ht}a_{22}^{(11)}}\tilde{\Phi}_s$, $\tilde{\epsilon} = \frac{\tilde{\xi}_{2211}^{(111)}}{\tilde{\xi}_{2121}^{(111)}}$, |
| $\delta\tilde{P}_3^{(i)}$ **** | $\delta\tilde{P}_{03}^{(1)} = \frac{\tilde{P}_{03}}{\cosh^2(\tilde{x}_1/w_{et})}$, $\tilde{P}_{03} = \frac{\chi\tilde{\Phi}_s}{w_{et}}$, <br> $\delta\tilde{P}_{03}^{(2)} = -\delta\tilde{P}_{03}^{(1)}$, $\delta\tilde{P}_{03}^{(3)} = \delta\tilde{P}_{03}^{(1)}$, <br> $\delta\tilde{P}_{03}^{(4)} = -\delta\tilde{P}_{03}^{(1)}$. | $\delta\tilde{P}_{03}^{(1)} = -\frac{2\mu\tilde{P}_{03}\tanh\left(\frac{\tilde{x}_1}{w_{ht}}\right)}{\cosh^2(\tilde{x}_1/w_{ht})}$, $\tilde{P}_{03} = \frac{\chi\tilde{\Phi}_s}{w_{ht}}$, <br> $\delta\tilde{P}_{03}^{(2)} = -\delta\tilde{P}_{03}^{(1)}$, $\delta\tilde{P}_{03}^{(3)} = \delta\tilde{P}_{03}^{(1)}$, <br> $\delta\tilde{P}_{03}^{(4)} = -\delta\tilde{P}_{03}^{(1)}$. |

*The rotated coordinates $\tilde{x}_1 = \frac{1}{\sqrt{2}}(x_1 + x_2)$, $\tilde{x}_2 = \frac{1}{\sqrt{2}}(x_2 - x_1)$ and $\tilde{x}_3 = x_3$ are introduced for the 90-degree TWs. The distance from the TW is equal to $\tilde{x}_1$. The tilts $\tilde{\Phi}_j^{(i)}(\vec{x})$ are sign-alternating functions of $\tilde{x}_2$ and $\tilde{x}_3$, but these functions are omitted in the table for the sake of brevity.

**We regard that $\xi_{jklm}^{(iii)} = \tilde{\xi}_{jklm}^{(111)}$ and $a_{jk}^{(ii)} = a_{jk}^{(11)}$, and so $\xi_{jklm}^{(iii)} = \tilde{\xi}_{jklm}^{(111)}$ due to the cubic symmetry of the parent phase.

***The vacuum permittivity $\varepsilon_0 = 8.85 \cdot 10^{-12}$F/m and the relative background permittivity $\varepsilon_b$ changes from 3 to 10, as a rule.

****The sublattice polarization components $\delta\tilde{P}_3^{(i)}(\vec{x})$ are sign-alternating functions of $\tilde{x}_2$ and $\tilde{x}_3$, but these functions are omitted in the table for the sake of brevity.

**Table II.** Order parameters near "easy" (i.e., lower energy) and "hard" (i.e., higher energy) 180-degree APBs in the orthorhombic phase of AFD ferroelastics with a cubic parent phase

| Component | Easy 180-degree APBs | Hard 180-degree APBs |
|---|---|---|
| $\Phi_1^{(i)*}$ | $\Phi_1^{(1)} = \Phi_s \tanh\left(\frac{x_1}{w_{ea}}\right)$, $\Phi_1^{(2)} = -\Phi_1^{(1)}$, $\Phi_1^{(3)} = \Phi_1^{(1)}$, $\Phi_1^{(4)} = -\Phi_1^{(1)}$ | $\Phi_1^{(1)} = \frac{\Phi_s}{\cosh(x_1/w_{ha})}$, $\Phi_1^{(2)} = -\Phi_1^{(1)}$, $\Phi_1^{(3)} = \Phi_1^{(1)}$, $\Phi_1^{(4)} = -\Phi_1^{(1)}$ |
| $\Phi_2^{(i)}$ | 0 | $\Phi_2^{(1)} = \Phi_s \tanh\left(\frac{x_1}{w_{ha}}\right)$, $\Phi_2^{(2)} = -\Phi_2^{(1)}$, $\Phi_2^{(3)} = \Phi_2^{(1)}$, $\Phi_2^{(4)} = -\Phi_2^{(1)}$ |
| $\Phi_3^{(i)}$ | 0 | 0 |
| $\delta P_1^{(i)**}$ | $\eta \Phi_s P_{01} \frac{\tanh\left(\frac{x_1}{w_{ea}}\right)}{\cosh^2(x_1/w_{ea})}$, $P_{01} = -\frac{\xi_{1111}^{(111)}}{w_{ea} a_{11}^{(11)}} \Phi_s$, $\eta = \frac{\varepsilon_0 \varepsilon_b a_{11}^{(11)}}{1+\varepsilon_0 \varepsilon_b a_{11}^{(11)}}$ | $\eta \Phi_s P_{01} \frac{\tanh\left(\frac{x_1}{w_{ha}}\right)}{\cosh^2(x_1/w_{ha})}$, $P_{01} = \frac{\xi_{1221}^{(111)} - \xi_{1111}^{(111)}}{w_{ha} a_{11}^{(11)}} \Phi_s$, $\eta = \frac{\varepsilon_0 \varepsilon_b a_{11}^{(11)}}{1+\varepsilon_0 \varepsilon_b a_{11}^{(11)}}$ |
| $\delta P_2^{(i)}$ | 0 | $\Phi_s P_{02} \frac{1-\epsilon \sinh^2(x_1/w_{ha})}{\cosh^3(x_1/w_{ha})}$, $P_{02} = -\frac{\xi_{2121}^{(111)}}{w_{ha} a_{22}^{(11)}} \Phi_s$, $\epsilon = \frac{\xi_{2211}^{(111)}}{\xi_{2121}^{(111)}}$ |
| $\delta P_3^{(i)**}$ | 0 | $\delta \tilde{P}_{03}^{(1)} = \frac{P_{03}}{\cosh^2(x_1/w_{ha})}$, $P_{03} = -\frac{\chi \Phi_s}{w_{ha}}$, $\delta \tilde{P}_{03}^{(2)} = -\delta \tilde{P}_{03}^{(1)}$, $\delta \tilde{P}_{03}^{(3)} = \delta \tilde{P}_{03}^{(1)}$, $\delta \tilde{P}_{03}^{(4)} = -\delta \tilde{P}_{03}^{(1)}$ |

*The tilts $\Phi_j^{(i)}(\vec{x})$ are sign-alternating functions of $x_2$ and $x_3$, but these functions are omitted in the table for the sake of brevity.

**The sublattice polarization components $\delta P_3^{(i)}(\vec{x})$ are sign-alternating functions of $x_2$ and $x_3$, but these functions are omitted in the table for the sake of brevity.

Distributions of the nonzero components of the normalized AFD order parameter (e.g., oxygen tilt) $\tilde{\Phi}_j^{(i)}/\tilde{\Phi}_s$ and polarization $\delta \tilde{P}_j^{(i)}/\tilde{P}_{0j}$ near easy and hard 90-degree TWs in the AFD ferroelastic are shown in **Fig. 2**. All physical quantities are listed in the rotated coordinates, $\tilde{x}_1 = \frac{1}{\sqrt{2}}(x_1 + x_2)$, $\tilde{x}_2 = \frac{1}{\sqrt{2}}(x_2 - x_1)$ and $\tilde{x}_3 = x_3$, which are commonly used for description of 90-degree TWs. The distance from the TW is equal to $\tilde{x}_1$, and the coordinate axes $\tilde{x}_2$ and $\tilde{x}_3$ are parallel to the wall plane in the rotated frame.

The components $\tilde{\Phi}_1^{(i)}$ and $\tilde{\Phi}_2^{(i)}$ are sign-alternating in a checkerboard order in neighboring sublattices (see **Table I**, **Fig. S1** [29], and color insets in **Figs. 2(a), 2(b), 2(d)** and **2(e)**). Because of a small positive parameter $\mu$, the envelopes of $\tilde{\Phi}_1^{(i)}$ are slightly varying functions of $\tilde{x}_1$ near easy TWs (see dotted red curves in **Fig. 2(a)**), and the envelopes of $\tilde{\Phi}_2^{(i)}$ are slightly varying functions of $\tilde{x}_1$ near hard TWs (see dotted green curves in **Figs. 2(e)**). The envelopes of $\tilde{\Phi}_1^{(i)}$ are even functions of $\tilde{x}_1$ near easy TWs (see dotted red curves in **Fig. 2(a)**), and the envelopes of $\tilde{\Phi}_2^{(i)}$ are even functions of $\tilde{x}_1$ near hard TWs (see dotted green curves in **Figs. 2(e)**). The envelopes of $\tilde{\Phi}_1^{(i)}$ are odd functions of $\tilde{x}_1$, which vanish at hard TWs (see dotted red curves in **Figs. 2(d)**), and the envelopes of the $\tilde{\Phi}_2^{(i)}$ are odd functions of $\tilde{x}_1$, which vanish at easy TWs (see dotted green curves in **Figs. 2(b)**). The component $\tilde{\Phi}_3^{(i)}$ is absent near the TWs.

The components $\delta\tilde{P}_1^{(i)}$ and $\delta\tilde{P}_2^{(i)}$ of sublattice polarization, which are induced by the linear-quadratic flexo-AFD coupling with the tilts $\tilde{\Phi}_1^{(i)}$ and $\tilde{\Phi}_2^{(i)}$, are not sign-alternating and proportional to $\eta\Phi_s^2$ and $\Phi_s^2$, respectively (see red and green curves in **Figs. 2(c)** and **2(f)**). However, the magnitudes of $\delta\tilde{P}_1^{(i)}$ and $\delta\tilde{P}_2^{(i)}$ are relatively small due to the following reasons. The absolute value $\left|\delta\tilde{P}_2^{(i)}\right| \leq 1-5$ μC/cm$^2$ for AFD perovskites like $SrTiO_3$, because the absolute value of $\Phi_s$ (e.g., oxygen octahedra tilt angle) is less than several degrees in these materials [3, 4]. The magnitude of $\delta\tilde{P}_1^{(i)}$ is negligibly small in comparison with $\delta\tilde{P}_2^{(i)}$ because the component $\delta\tilde{P}_1^{(i)}$, that is normal to the plane of the TWs, is strongly suppressed by the depolarization field (see red curves in **Figs. 2(c)** and **2(f)**). According to **Table I**, the ratio $\frac{\delta\tilde{P}_2^{(i)}}{\delta\tilde{P}_1^{(i)}}$ is proportional to the very small factor $\eta = \frac{\varepsilon_0\varepsilon_b a_{11}^{(ii)}}{1+\varepsilon_0\varepsilon_b a_{11}^{(ii)}}$, that is less than $10^{-2}$, since the background permittivity $\varepsilon_b$ is less than 10, $\left|a_{11}^{(ii)}\right| \leq 10^8$ m/F and therefore $\eta < \left|\varepsilon_0\varepsilon_b a_{11}^{(ii)}\right| \leq 0.01$. The component $\delta\tilde{P}_2^{(i)}$ is even function of $\tilde{x}_1$ that reaches maximum at the TWs; and $\delta\tilde{P}_1^{(i)}$ is an odd function of $\tilde{x}_1$ that is zero at the TWs (see red and green curves in **Figs. 2(c)** and **2(f)**).

The sublattice polarization component $\delta\tilde{P}_3^{(i)}$ is sign-alternating in a checkerboard order near the 90-degree TWs and proportional to $\chi\Phi_s$, because it is induced by the bilinear flexo-AFD coupling (see solid blue curves and color insets in in **Figs. 2(c)** and **2(f)**). The envelopes of $\delta\tilde{P}_3^{(i)}$ are even functions of $\tilde{x}_1$ near easy TWs (see dotted blues curves in **Fig. 2(c)**) and odd functions of $\tilde{x}_1$ near hard TWs (see blues curves in **Fig. 2(f)**). The amplitude of $\delta\tilde{P}_3^{(i)}$ can be significantly larger than $\left|\delta\tilde{P}_1^{(i)}\right|$ and $\left|\delta\tilde{P}_2^{(i)}\right|$ (see dashed blue curves in **Figs. 2(c)** and **2(f)**). For typical material parameters (e.g., for $\Phi_s = 6°$, $\tilde{\rho} = 2$ and $\tilde{\varepsilon} = -0.5$) the amplitude of $\delta\tilde{P}_3^{(i)}$ is more than 10 times larger than the maximal value of $\left|\delta\tilde{P}_2^{(i)}\right|$ near easy TWs (see **Fig. 2(c)**). The amplitude of $\delta\tilde{P}_3^{(i)}$ is more than 3 times larger than the maximal value of $\left|\delta\tilde{P}_2^{(i)}\right|$ near hard TWs (see **Fig. 2(f)**). The amplitude of $\delta\tilde{P}_3^{(i)}$ is more than $10^2 - 10^3$ times larger than the maximal value of $\left|\delta\tilde{P}_1^{(i)}\right|$ near TWs. Notably that the linear-quadratic flexo-AFD coupling cannot induce the component $\delta\tilde{P}_3^{(i)}$ at the TWs.

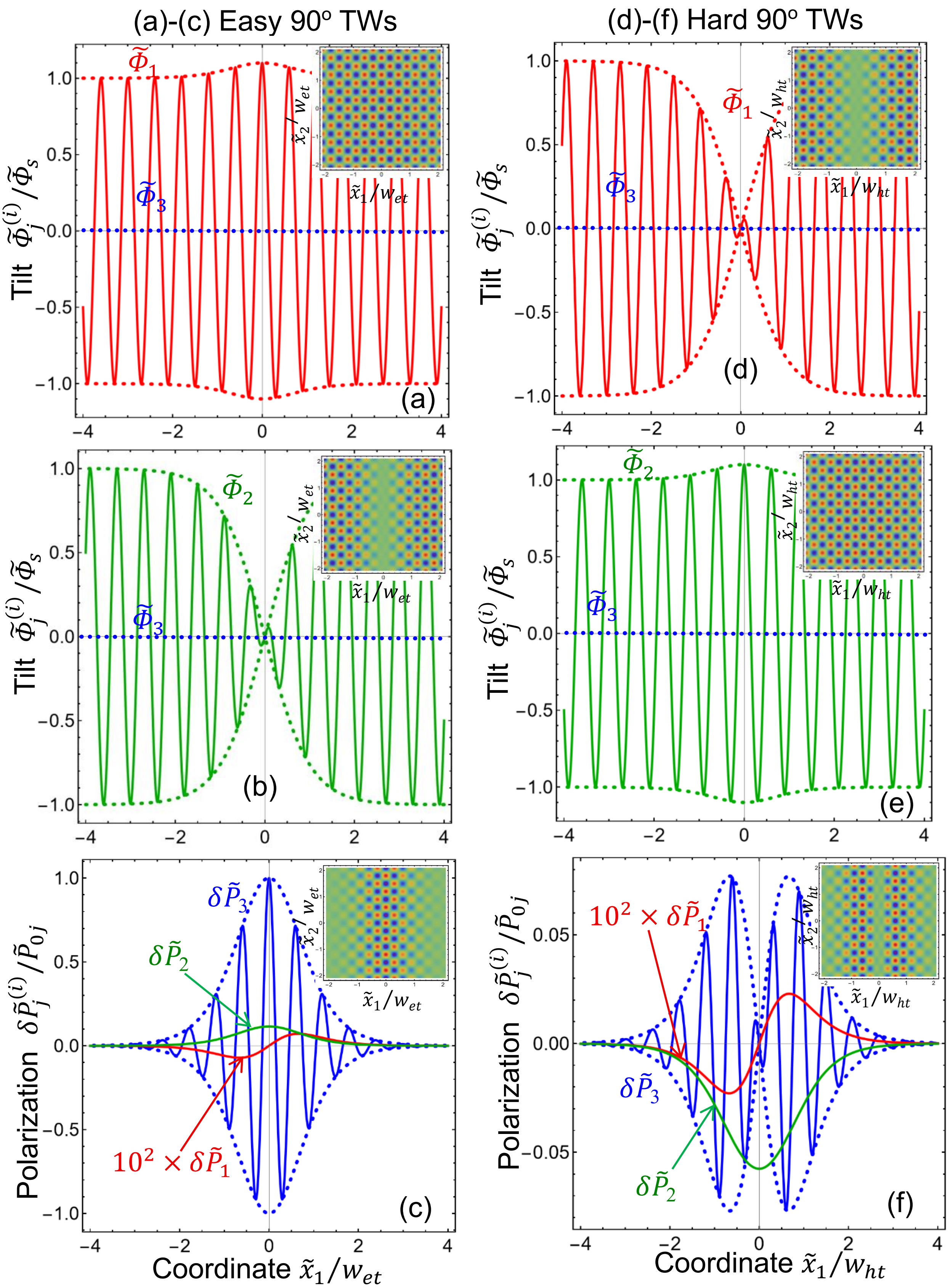

**FIGURE 2.** Distributions of the nonzero components of the normalized tilt $\tilde{\Phi}_j^{(i)}/\tilde{\Phi}_s$ **(a, b, d, e)** and polarization $\delta\tilde{P}_j^{(i)}/\tilde{P}_{0j}$ **(e, f)** components near easy **(a, b, c)** and hard **(b, d, f)** 90-degree TWs in the AFD ferroelastic. The distance from the TW is equal to $\tilde{x}_1$, the coordinate $\tilde{x}_2 = 0$. The envelope functions are shown by dotted curves.

The amplitude of $\delta\tilde{P}_1^{(i)}/\tilde{P}_{01}$ is multiplied by factor $10^2$. Color maps of the tilt components $\tilde{\Phi}_1^{(i)}$ and $\tilde{\Phi}_2^{(i)}$, and polarization component $\delta\tilde{P}_3^{(i)}$, which are checkerboard-type sign-alternating functions of $\tilde{x}_1$ and $\tilde{x}_2$, are shown in insets. The plots are calculated using **Table I** for parameters $\Phi_S = 6°$, $\mu = 0.1$, $\eta = 0.01$, $\tilde{\rho} = 2$ and $\tilde{\epsilon} = -0.5$.

Distributions of the nonzero components of the normalized tilt $\Phi_j^{(i)}/\Phi_S$ and polarization $\delta P_j^{(i)}/P_{0j}$ near easy and hard 180-degree APBs in the AFD ferroelastic are shown in **Fig. 3**. The distance from the APB is equal to $x_1$, and the coordinate axes $x_2$ and $x_3$ are parallel to the APB plane. The component $\Phi_1^{(i)}$ is sign-alternating in the checkerboard order in neighboring sublattices (see color insets in **Figs. 3(a)** and **3(d)**, and **Fig. S1** [29]). The envelopes of $\Phi_1^{(i)}$ are odd functions of $x_1$ near easy APBs (see dotted curves in **Fig. 3(d)**) and even functions of $x_1$ near hard APBs (see dotted curves in **Fig. 3(a)**). The components $\Phi_2^{(i)}$ and $\Phi_3^{(i)}$ are absent near easy APBs. The sign-alternating component $\Phi_2^{(i)}$ has envelopes, which are odd functions of $x_1$ near hard APBs (see color inset and dotted curves in **Fig. 3(e)**). The component $\Phi_3^{(i)}$ is absent near hard APBs.

The components of sublattice polarization $\delta P_1^{(i)}$ and $\delta P_2^{(i)}$, induced by the linear-quadratic flexo-AFD coupling with the tilts $\Phi_1^{(i)}$ and $\Phi_2^{(i)}$, are not sign-alternating and small being proportional to $\eta\Phi_S^2$ and $\Phi_S^2$, respectively (see the red curve in **Fig. 3(c)**). The magnitude of $\delta P_1^{(i)}$, that is normal to the plane of the easy APB, is negligibly small due to the strong depolarization field. The components $\delta P_2^{(i)}$ and $\delta P_3^{(i)}$ are absent near easy APBs.

The magnitudes of $\delta P_1^{(i)}$ and $\delta P_2^{(i)}$ are relatively small near hard APBs due to the same reasons as in the case of TWs. The magnitude of $\delta P_2^{(i)}$ does not exceed 1 – 3μC/cm$^2$ for AFD perovskites like $SrTiO_3$, because the absolute value of the $\Phi_S$ is less than several degrees [3, 4].

The sublattice polarization component $\delta P_3^{(i)}$ is sign-alternating in a checkerboard order near the hard 180-degree APBs and proportional to $\chi\Phi_S$, because it is induced by the bilinear flexo-AFD coupling (see the solid blue curve and color inset in **Fig. 3(f)**). The envelopes of $\delta P_3^{(i)}$ are even functions of $\tilde{x}_1$ near the APB (see dotted blues curves in **Fig. 3(f)**). For typical material parameters (e.g., for $\Phi_S = 6°$, $\eta = 0.01$ and $\epsilon = -0.5$) the amplitude of $\delta P_3^{(i)}$ is about 10 times larger than the maximal value of $\left|\delta P_2^{(i)}\right|$ and more than $10^3$ times larger than the maximal value of $\left|\delta P_1^{(i)}\right|$ near hard APBs (compare dotted blue and green curves in **Fig. 3(f)**). Notably that the linear-quadratic flexo-AFD coupling cannot induce the component $\delta P_3^{(i)}$ at hard APBs. The component $\delta P_3^{(i)}$ is absent at easy APBs, because the components $\Phi_2^{(i)}$ and $\Phi_3^{(i)}$ are absent, and $\nabla \times \vec{\Phi}^{(i)} = 0$ near easy APBs.

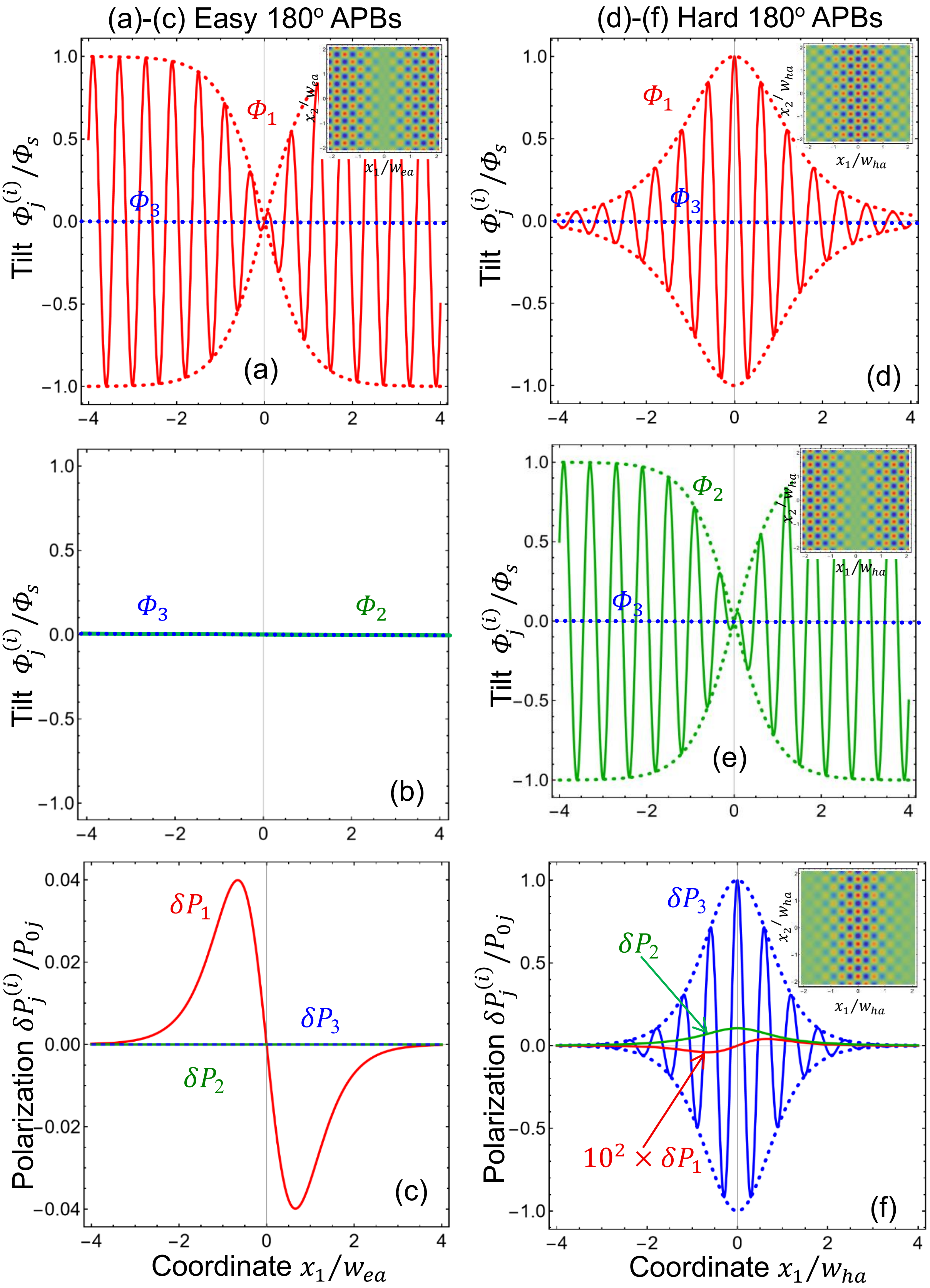

**FIGURE 3.** Distributions of the nonzero components of the normalized tilt $\Phi_j^{(i)}/\Phi_S$ **(a, b, d, e)** and polarization $\delta P_j^{(i)}/P_{0j}$ **(e, f)** near easy **(a, b, c)** and hard **(b, d, f)** 180-degree APBs in the AFD ferroelastic. The envelope

functions are shown by dotted curves. The distance from the APB is equal to $x_1$, the coordinate $x_2 = 0$. The amplitude of $\delta P_1^{(i)}/P_{01}$ is multiplied by factor $10^2$. Color maps of tilt components $\tilde{\Phi}_1^{(i)}$ and $\tilde{\Phi}_2^{(i)}$, and polarization component $\delta \tilde{P}_3^{(i)}$, which are checkerboard-type sign-alternating functions of $x_1$ and $x_2$, are shown in insets. The plots are calculated using **Table II** for parameters $\Phi_s = 6°$, $\eta = 0.01$ and $\epsilon = -0.5$.

## IV. Local band bending near TWs and APBs in AFD ferroelastics

Considering the influence of Bardeen-Shockley deformation potential theory [30], elastic strains induced by the flexo-AFD couplings can produce the local band bending near the TWs and APBs in the AFD ferroelastics [7]. Within the deformation potential theory, the local changes of the band structure induced by anisotropic strains are determined by the Bir and Pikus perturbation approach [31]. The most general form of the Bir-Pikus Hamiltonian is

$$\hat{H}_{PB}(\vec{x}) = \hat{D}_{ij}^{PB} u_{ij}^{f}(\vec{x}), \tag{6}$$

where $\hat{D}_{ij}^{PB}$ is deformation potential operator; $u_{ij}^{f}(\vec{x})$ is the strain variation, that is spatially inhomogeneous in the considered case of the near-wall acoustic phonons. The dispersion laws of the holes in the the valence band, $E_V(\vec{k})$, and electrons in the conduction band, $E_C(\vec{k})$, can be estimated using the effective mass approximation and perturbation approach. Within the approach, the Bir-Pikus Hamiltonian $\hat{H}_{PB}(\vec{x})$ is considered as a perturbation to e.g., the Kohn-Luttinger [32] Hamiltonian $\hat{H}_{KL}(\vec{x})$ listed in **Supplement S2** [29]. The dispersion laws are rederived in **Supplement S2** [29]. They are well known and have the form:

$$E_V(\vec{k}) = E_{V0} + S_{Vk} \pm \sqrt{T_{Vk}^2 + |U_{Vk}^2| + |W_{Vk}^2|}, \tag{7a}$$

$$E_C(\vec{k}) = E_{C0} - S_{Ck} \mp \sqrt{T_{Ck}^2 + |U_{Ck}^2| + |W_{Ck}^2|}. \tag{7b}$$

The values $E_{V0}$ and $E_{C0}$ are energetic positions of the bands far from the TWs and APBs. The sign "+" or "−" before the radical corresponds to the "light" and "heavy" holes (or electrons). The k-dependent and strain-dependent parameters $S_{Bk}$, $T_{Bk}$, $U_{Bk}$ and $W_{Bk}$, where $B = V$ for holes or $B = C$ for electrons, are listed in **Supplement S2** [29].

In a particular case $k_2 = 0$, considering nonzero flexo-AFD strain $u_{13}^{f} \neq 0$ and assuming that other strain components are negligibly small (e.g., due to the small electrostriction coupling), the dispersion laws Eqs. (7a)-(7b) acquire a relatively simple form

$$E_V(\vec{k}) \approx E_{V0} - \frac{\hbar^2}{2m_0}\gamma_1^p (k_1^2 + k_3^2) \mp \frac{\sqrt{3}\hbar^2}{m_0}\gamma_3^p \left|k_1 k_3 - D_f^p u_{13}^k\right|, \tag{7c}$$

$$E_C(\vec{k}) \approx E_{C0} + \frac{\hbar^2}{2m_0}\gamma_1^n (k_1^2 + k_3^2) \pm \frac{\sqrt{3}\hbar^2}{m_0}\gamma_3^n \left|k_1 k_3 - D_f^n u_{13}^k\right|. \tag{7d}$$

where $m_0$ is the mass of free electron, $\gamma_1^{n,p}$ and $\gamma_3^{n,p}$ are Luttinger parameters determining effective masses, the parameters $D_f^p = \frac{m_0 d}{\sqrt{3}\hbar^2 \gamma_3^p}$ and $D_f^n = \frac{m_0 d}{\sqrt{3}\hbar^2 \gamma_3^n}$ are proportional to the shear deformation potential constant $d$, the matrix element $u_{13}^k = \langle \psi_k | \hat{u}_{ij}^f | \psi_k \rangle \sim P_{03} \sim \chi \Phi_s$ changes its sign with the sign of polarization amplitude $P_{03}$. Other parameters and details of Eqs. (7) derivation are presented in **Supplement S2** [29]. Expressions (7c)-(7d) are derived for small $k$ corresponding to the parabolic (or harmonic) approximation valid near the bottom of conduction band (or near the top of valence band). The simplest way to consider anharmonicity and to increase k-limits of Eqs. (7) applicability is to make a substitution, $k_1 \to k_0 \sin\left(\frac{k_1}{k_0}\right)$ and $k_3 \to k_0 \sin\left(\frac{k_3}{k_0}\right)$, where $k_0$ is the inverse lattice vector in the corresponding direction of the Brillouin space.

Color maps of the local band bending $E_C(\vec{k})$, calculated at the structural walls in the AFD ferroelastics with prevailing bilinear flexo-AFD coupling for positive ("dipole up") and negative ("dipole down") polarization $P_{03}$ of the corresponding sub-lattice, are shown in **Figs. 5(a)** and **5(b)**, respectively (see **Table II** for $P_{03}$ definition). It is seen from these figures that the sub-bands have alternating profiles along the perpendicular (diagonal) directions $\tilde{q}_K = (11)$ (namely $k_3 = k_1$) and $\tilde{q}_{K'} = (1\bar{1})$ (namely $k_1 = -k_3$) under the change $P_{03} \to -P_{03}$ (compare positions of the light-blue and dark-blue wells, located in the $K$ and $K'$ points of the first Brillouin zone, in **Figs. 5(a)** and **5(b)**). The effect of alternating bands, emerging at $P_{03} \to -P_{03}$, is illustrated explicitly in **Figs. 5(c)** and **5(d)**. Here, the dispersion curves, $E_C(\vec{k})$ and $E_V(\vec{k})$, are calculated for $k_3 = k_1$ and $k_1 = -k_3$, and $k_2 = 0$ (for both plots). Red and blue curves correspond to the positive ("dipole up") and negative ("dipole down") polarization $P_{03}$ in each sub-band. Solid and dashed curves correspond to the dispersion curves of light and heavy carriers, respectively. The rotation at 90-degrees is dictated by the atomic structure shown in **Fig. 1(d).**

Hence, we can conclude that the bilinear flexo-AFD coupling leads to the alternating-type splitting of the carrier sub-bands with respect to the sign of emerging polarization $\delta P_{03}^{(m)}$ near the structural domain boundaries. We also have shown that under specific choice of parameters the splitting is of Rashba-Edelstein type (see **Fig. S2** in **Supplement S2** [29]).

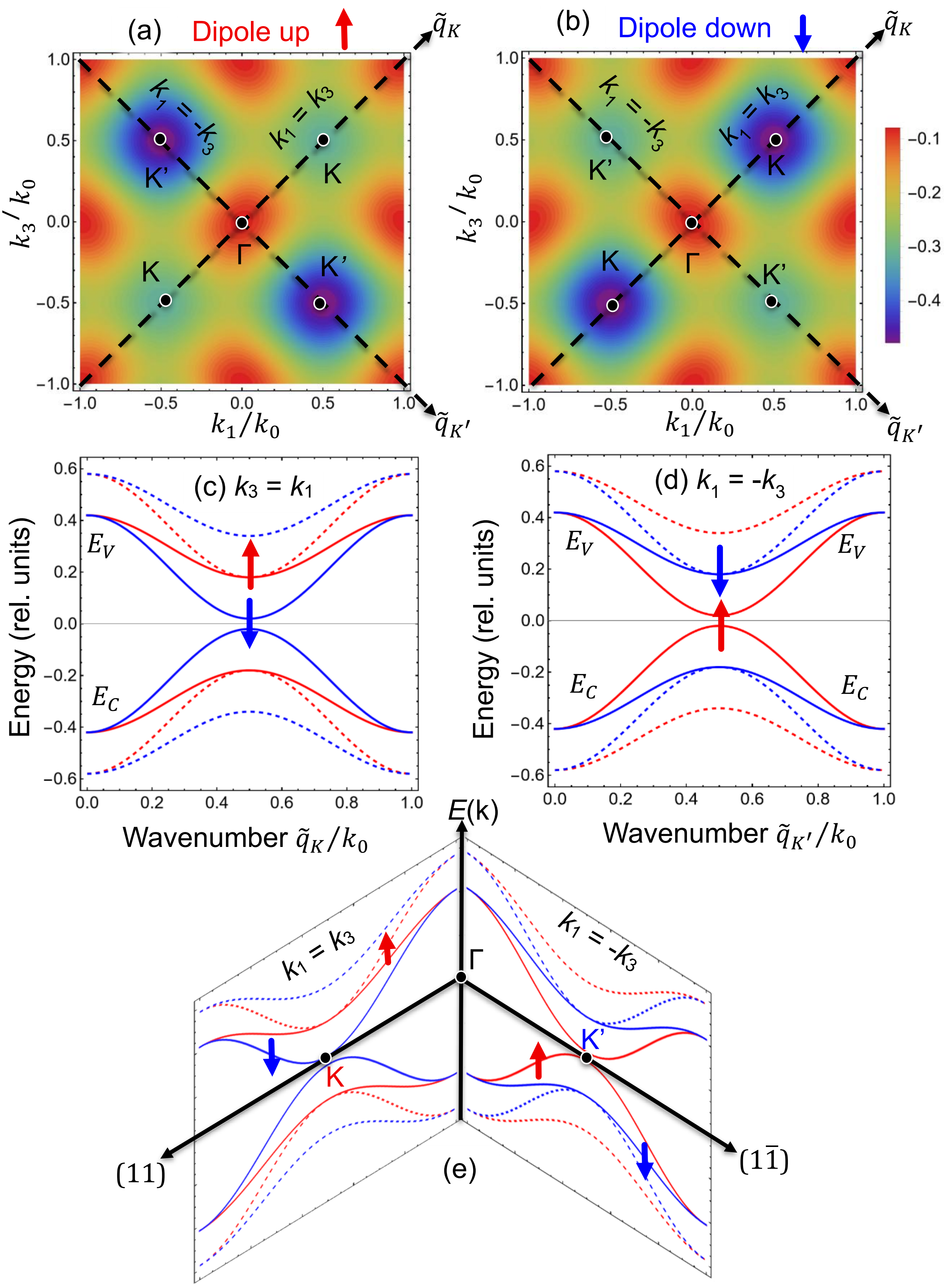

**FIGURE 4.** Color maps of the local band bending $E_C(\vec{k})$ (in normalized units), calculated at the structural walls in the AFD ferroelastics with prevailing bilinear flexo-AFD coupling for positive ("dipole up", part **(a)**) and negative ("dipole down", part **(b)**) polarization $P_{03}$ of the corresponding sub-lattice, $k_2 = 0$. Dispersion

curves $E_C(\vec{k})$ and $E_V(\vec{k})$ calculated for $k_3 = k_1$ **(c)** and $k_1 = -k_3$ **(d)**; $k_2 = 0$ for both plots. The coordinates $\tilde{q}_K$ and $\tilde{q}_{K'}$ correspond to the perpendicular directions $k_3 = k_1$ and $k_1 = -k_3$, respectively. The directions $\tilde{q}_K$ and $\tilde{q}_{K'}$, which are rotated at 90-degree with respect to $k_1$ and $k_3$ coordinate system, are shown by thick dashed lines with ending arrows in panels (a) and (b). The parameter $u_{13}^k \sim P_{03}$ is regarded as $k$-independent. Other parameters are listed in **Supplement S2** [29]. Red and blue curves correspond to the positive ("dipole up") and negative ("dipole down") polarization $P_{03}$ in the sub-bands. Solid and dashed curves correspond to light and heavy carriers, respectively. The parameter $k_0$ is the inverse lattice vector in corresponding direction of the Brillouin space. **(e)** Alternating bands in the 90-degree cross-section of the Brillouin space.

The strain variation is $u_{ij}^f(\vec{x}) = u_{ij}(x_1, x_2, x_3) - u_{ij}^s(\infty, x_2, x_3)$; $u_{ij}^s$ is the spontaneous strain far from the wall, where $\vec{\Phi}^2 \to \Phi_S^2$, $\vec{P}^2 \to 0$ and $\sigma_{ij} \to 0$. Relation between elastic stresses $\sigma_{ij}$ and strains $u_{ij}$ follows from minimization of the free energy with respect to the strain, $\sigma_{ij} = \frac{\partial G}{\partial u_{ij}}$, and has the form:

$$\sigma_{mn} = c_{mnkl} u_{kl} - \tilde{R}_{mnkl}^{(ij)} \Phi_k^{(i)} \Phi_l^{(j)} - \tilde{Q}_{mnkl}^{(ij)} P_k^{(i)} P_l^{(j)} + f_{mnkl}^{(i)} \frac{\partial P_k^{(i)}}{\partial x_l}. \tag{8}$$

Since the stress and polarization vanish far from the wall, the strain variation $u_{ij}^f$ and the spontaneous strain $u_{ij}^s$ can be estimated from expression (8) as:

$$u_{ij}^f = s_{ijkl}\sigma_{kl} + R_{ijkl}^{(mn)}\left(\Phi_k^{(m)}\Phi_l^{(n)} - \Phi_{ks}^{(m)}\Phi_{ls}^{(n)}\right) + Q_{ijkl}^{(mn)} P_k^{(m)} P_l^{(n)} - F_{ijkl}^{(m)} \frac{\partial P_k^{(m)}}{\partial x_l}, \tag{9a}$$

$$u_{ij}^s = R_{ijkl}^{(mn)} \Phi_{ks}^{(m)} \Phi_{ls}^{(n)}, \tag{9b}$$

where $s_{ijkl}$ are elastic compliances, $R_{ijkl}^{(mn)} = c_{rsij}^{-1}\tilde{R}_{rskl}^{(mn)}$, $Q_{ijkl}^{(mn)} = c_{rsij}^{-1}\tilde{Q}_{rskl}^{(mn)}$ and $F_{ijkl}^{(m)} = c_{rsij}^{-1} f_{rskl}^{(m)}$. The first term in the right-hand side of Eq.(9a) corresponds to the linear Hooke's law, the second term is proportional to $\Phi_S^2$, not sign-alternating and determined by the AFD ordering only. The third term in the right-hand side of Eq.(9a) is also proportional to $\Phi_S^2$ and not sign-alternating. The term contains the contribution of the HLS coupling (proportional to $\eta^2\Phi_S^2$), the linear-quadratic flexo-AFD coupling (proportional to $\zeta^2\Phi_S^2$), and the bilinear flexo-AFD coupling (proportional to the $\chi^2\Phi_S^2$), as well as their cross-terms. As a rule, all terms proportional to $\Phi_S^2$ are small, because the absolute value of the spontaneous tilt $\Phi_S$ is less than several degrees [3, 4]. The last term on the right-hand side of Eq.(9a) contains the sign-alternating checkerboard type contribution proportional to the product $f \cdot \chi\Phi_S$ coming from the bilinear flexo-AFD coupling only. The last term may dominate in alterelectric-type materials. Typical distributions of the polarization component $\delta P_3^{(i)}$ and the local strain $\delta u_{13}^f$ as a function of transverse coordinates calculated near TWs and APBs in the AFD ferroelastics with prevailing bilinear flexo-AFD coupling are shown in **Fig. 5**.

To summarize, the local changes of the band structure, induced by the HLS, linear-quadratic and bilinear flexo-AFD couplings, should differ significantly according to Eqs.(6) and (9). Namely, the local band bending is proportional to the $\Phi_s^2$ for prevailing HLS and/or linear-quadratic flexo-AFD couplings, or proportional to the first power of $\Phi_s$ for dominating bilinear flexo-AFD coupling.

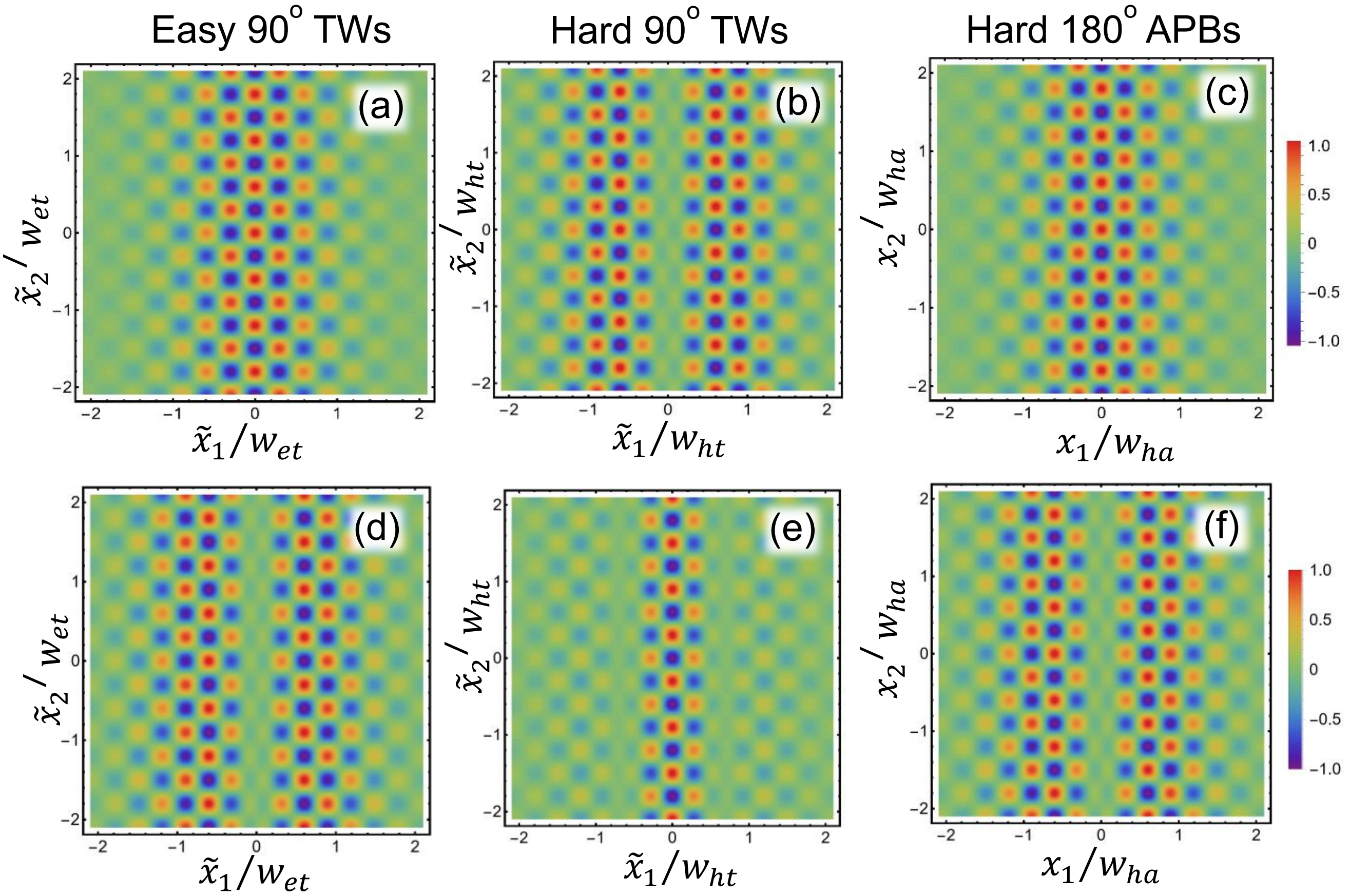

**FIGURE 5.** Distributions of the polarization component $\delta P_3^{(i)}$ **(a-c)** and the local strain $\delta u_{13}^{f}$ **(d-f)** as a function of transverse coordinates calculated near easy **(a, d)** and hard **(b, e)** TWs, and hard APBs **(c, f)** in the AFD ferroelastics with prevailing bilinear flexo-AFD coupling. Color scales are given in normalized units. Parameters are the same as in **Fig. 2** and **3**.

In addition to the strain contribution via the deformation potential mechanism, the depolarization electric field $\delta E_1^d$, related with the polarization component $\delta P_1^{(i)}$ as $\delta E_1^d \cong -\frac{\delta P_1^{(i)}}{\varepsilon_0 \varepsilon_b}$, contributes to the local band bending. Despite $\delta P_1^{(i)}$ is very small, the depolarization field can be significant, because of the large factor $\frac{1}{\varepsilon_0 \varepsilon_b} \geq 10^{10}$ m/F. Thus, even a tiny polarization component normal to the plane of the TW or APB can induce a strong depolarization field.

The strain-induced variations of band structure and the depolarization field induce local changes of the electrochemical potentials, which influence the concentration of free electrons $n(\vec{x})$ and holes $p(\vec{x})$ near the TWs and APBs. The concentration changes can be estimated in the Boltzmann approximation as:

$$n(\vec{x}) \approx n_0 \exp\left[\frac{1}{k_B T}\left(D^C_{ij} u^f_{ij}(\vec{x}) + e\int_{-\infty}^{x_1} \delta E^d_1\, dx\right)\right], \tag{10a}$$

$$p(\vec{x}) \approx p_0 \exp\left[\frac{-1}{k_B T}\left(D^V_{ij} u^f_{ij}(\vec{x}) + e\int_{-\infty}^{x_1} \delta E^d_1\, dx\right)\right]. \tag{10b}$$

Here $k_B = 1.3807\times10^{-23}$ J/K is the Boltzmann constant, $T$ is the absolute temperature, and $e = 1.6\times10^{-19}$ C is the electron charge. The terms proportional to $D^C_{ij} u^f_{ij}(\vec{x})$ and $D^V_{ij} u^f_{ij}(\vec{x})$ are the contribution of elastic strains, and the terms proportional to the integral $\delta\varphi = -\int_{-\infty}^{x_1} \delta E^d_1\, dx$ are the contribution of the "net" depolarization field to the free carrier redistribution at the TWs and APBs in AFD ferroelastics. Due to the self-screening by free carriers, the screened or "dressed" depolarization field reduces strongly, and the terms proportional to $\frac{e\delta\varphi(\vec{x})}{k_B T}$ can become small enough near the TWs and APBs.

## IV. Discussion and conclusions

Using the four sublattices model and Landau-Ginsburg-Devonshire approach, we show that the linear gradient-type coupling between the sublattice polarization vector $\vec{P}^{(i)}$ and antiferrodistortive long-range order parameter – pseudovector $\vec{\Phi}^{(i)}$, that has the simplest form of Lifshitz invariant $\frac{\lambda^{(ii)}}{2}\left(\vec{P}^{(i)}\cdot\nabla\times\vec{\Phi}^{(i)} - \vec{\Phi}^{(i)}\cdot\nabla\times\vec{P}^{(i)}\right)$ and named "bilinear flexo-AFD coupling", can emerge in all AFD ferroelastics, since it is symmetry-allowed. We reveal that the bilinear flexo-AFD coupling can induce the sublattice-sensitive polarization at the 90-degree TWs and 180-degree APBs in AFD ferroelastics without any ferroelectric or antiferroelectric ordering. Since the induced polarization $\overrightarrow{\delta P}^{(i)}$ is perpendicular to $\vec{\Phi}^{(i)}$ and counter-directed in neighboring sublattices with checkerboard-type direction of $\vec{\Phi}^{(i)}$, such structure of $\overrightarrow{\delta P}^{(i)}$ corresponds to the alterelectric-type quadrupolar electric order (see insets in **Figs. 2(c), 2(f)** and **3(f)**). Since a common flexoelectric and/or linear-quadratic flexo-AFD couplings cannot induce the alterelectric-type polarization $\overrightarrow{\delta P}^{(i)} \perp \vec{\Phi}^{(i)}$ in AFD ferroelastics, we believe that the bilinear flexo-AFD coupling can be a reason for alterelectric long-range order at structural domain boundaries, surfaces or interfaces of AFD ferroelastics.

At the same time physical manifestations of the bilinear flexo-AFD coupling remain invisible in most nanostructured AFD ferroelectrics and antiferroelectrics due to the domination of piezoelectric and/or linear flexoelectric couplings. Nevertheless, obtained results can be verified experimentally, e.g., by the second harmonic generation (SHG), which intensity is proportional to the second power of $\overrightarrow{\delta P}^{(i)}$. One can expect the "bright" structural domain walls and antiphase boundaries in the SHG imaging of AFD ferroelastics without any ferroelectric or antiferroelectric ordering, but with alterelectric-type polar order (e.g., the component $\delta P_3^{(i)}$ considered in this work). The local PFM response, linearly proportional to the $\overrightarrow{\delta P}^{(i)}$, can reveal only the contributions from linear-biquadratic coupling, which are relatively small but not sigh-alternating (e.g., the component $\delta P_2^{(i)}$). Due to the deformation potential, elastic strains induced by the flexo-AFD couplings can change locally the band structure near the TWs and APBs in the AFD ferroelastics [7]. However, local changes of the band structure, induced by the HLS, bilinear and linear-quadratic flexo-AFD couplings, differ significantly. The band bending is linearly proportional to the AFD tilt, and the alternating-type splitting of corresponding sub-bands appears in the case of dominating bilinear flexo-AFD coupling. The splitting of sub-bands is absent, and the band bending is proportional to the second power of the AFD tilt in the case of prevailing HLS and/or linear-quadratic flexo-AFD couplings.

**Acknowledgements.** The work is primary supported as part of the Computational Materials Sciences Program funded by the US Department of Energy, Office of Science, Basic Energy Sciences, under Award Number DE-SC0020145 (A.N.M., E.A.E., A.S. and V.G.). The part of A.N.M. efforts (namely the discussion of obtained results) is also partially supported by National Academy of Sciences of Ukraine (grants No. 5.8/26-П and 1.4.В/222). The part of E.A.E. efforts (namely the supplemental numerical analysis) is also partially funded by National Academy of Sciences of Ukraine (grant III-6-26). V.G. also received support from the Center for 3D Ferroelectric Microelectronics Manufacturing (3DFeM2), an Energy Frontier Research Center funded by the U.S. Department of Energy, Office of Science, Office of Basic Energy Sciences Energy Frontier Research Centers program under Award Number DE-SC0021118. Results were visualized in Mathematica 15.0 [33]. E.A.E. and A.N.M. and gratefully acknowledge clarifying explanations of Acad. Vadim Loktev regarding the altermagnetism physical nature.

**Authors' contribution.** E.A.E. and A.N.M. contributed equally to the work. A.N.M. generated the research idea, formulated the problem, performed basic calculations, analyzed obtained results and wrote the manuscript draft. E.A.E. performed symmetry analysis, wrote the codes and performed numerical modeling. A.S. and V.G. worked on results explanation and manuscript improvement.

## SUPPLEMENTARY MATERIALS

### Supplement S1. Symmetry analysis of the LGD free energy

The bulk part of LGD free energy consists of the following contributions:

$$G = G_{EP} + G_{AFD} + G_{HLS} + G_{EL} + G_{ST} + G_{GR} + G_{FLR} - \frac{P_i E_i^d}{2} \qquad \text{(S.1)}$$

Here $E_i^d = -\partial\phi/\partial x_i$ are the components of depolarization field. We assume that bulk material paraelectric phase has cubic symmetry. The polarization-dependent density $\Delta G_{ferro}$ can be written as a Taylor series expansion of the components of polarization $P_i$, and tilts $\Phi_i$ as:

$$G_{AFD} = \sum_{k,l,i,j} b_{jl}^{(ik)} \Phi_j^{(i)} \Phi_l^{(k)} + \sum_{k,l,i,j,q,r,m,n} b_{jlqr}^{(ikmn)} \Phi_j^{(i)} \Phi_l^{(k)} \Phi_q^{(m)} \Phi_r^{(n)}, \qquad \text{(S.2)}$$

$$G_{EP} = \sum_{k,l,i,j} a_{jl}^{(ik)} P_j^{(i)} P_l^{(k)} + \sum_{k,l,i,j} a_{jlqr}^{(ikmn)} P_j^{(i)} P_l^{(k)} P_q^{(m)} P_r^{(n)}, \qquad \text{(S.3)}$$

$$G_{HLS} = \sum_{k,l,i,j,q,r,m,n} \eta_{jlqr}^{(ikmn)} \Phi_j^{(i)} \Phi_l^{(k)} P_q^{(m)} P_r^{(n)}. \qquad \text{(S.4)}$$

Electrostriction contribution to the LGD free energy is

$$G_{ST} = -\sum_{k,l,i,j} \tilde{Q}_{mnjl}^{(ik)} u_{mn} P_j^{(i)} P_l^{(k)} - \sum_{k,l,i,j} \tilde{R}_{mnjl}^{(ik)} u_{mn} \Phi_j^{(i)} \Phi_l^{(k)} \qquad \text{(S.5)}$$

Elastic energy is

$$G_{\mathrm{EL}} = \frac{1}{2} c_{ijkl} u_{ij} u_{kl} \qquad \text{(S.6)}$$

Gradient contribution (Ginzburg term) has a general form:

$$G_{GR} = \sum_{k,lq,r,i,j} \left( v_{klqr}^{(ij)} \frac{\partial \Phi_k^{(i)}}{\partial x_l} \frac{\partial \Phi_q^{(j)}}{\partial x_r} + g_{klqr}^{(ij)} \frac{\partial P_k^{(i)}}{\partial x_l} \frac{\partial P_q^{(j)}}{\partial x_r} \right) \qquad \text{(S.7)}$$

Flexoelectric and flexo-roto-electric couplings are represented as

$$G_{FLR} = \sum_{i=1}^{4} \left[ f_{mjkl}^{(i)} u_{mj} \frac{\partial P_k^{(i)}}{\partial x_l} + \xi_{mjkl} P_m^{(i)} \Phi_j^{(i)} \frac{\partial \Phi_k^{(i)}}{\partial x_l} \right] + \sum_{i,j} \lambda^{(ij)} \frac{\varepsilon_{qkl}}{2} \left( P_q^{(i)} \frac{\partial \Phi_k^{(j)}}{\partial x_l} - \Phi_k^{(j)} \frac{\partial P_q^{(i)}}{\partial x_l} \right). \qquad \text{(S.8)}$$

The summation in Eqs.(S.2) – (S.8) is performed with lower "space" indices (i.e. for $k, l, q, r = x, y, z$) and with upper "sublattice" indices (i.e. $i, j, m, n = 1, 2, 3, 4$). Nonzero components of Levi-Civita pseudo-tensor $\varepsilon_{qkl}$ are $\varepsilon_{123} = \varepsilon_{312} = \varepsilon_{231} = -\varepsilon_{321} = -\varepsilon_{132} = -\varepsilon_{213} = 1$ (other components are zero).

Equations of state for polarizations of sublattices satisfy the Euler-Lagrange equations,

$$\frac{\partial G}{\partial P_j^{(i)}} - \frac{\partial}{\partial x_k} \frac{\partial G}{\partial P_{j,k}^{(i)}} = 0\,, \qquad \text{(S.9a)}$$

which have the following form

$$2a_{jl}^{(ik)} P_l^{(k)} + 4a_{jlqr}^{(ikmn)} P_l^{(k)} P_q^{(m)} P_r^{(n)} + 2\eta_{rlqj}^{(nkmi)} \Phi_j^{(i)} \Phi_l^{(k)} P_q^{(m)} - 2\tilde{Q}_{mnjl}^{(ik)} u_{mn} P_l^{(k)} - f_{mkjl}^{(i)} \frac{\partial u_{mk}}{\partial x_l} +$$
$$\lambda^{(im)} \varepsilon_{jkl} \frac{\partial \Phi_k^{(m)}}{\partial x_l} + \xi_{jmnq}^{(ilk)} \Phi_m^{(l)} \frac{\partial \Phi_n^{(k)}}{\partial x_q} - g_{jlqr}^{(ik)} \frac{\partial^2 P_q^{(k)}}{\partial x_l \partial x_r} = 0. \qquad \text{(S.9b)}$$

Equations of state for AFD parameters of sublattices satisfy the Euler-Lagrange equations,

$$\frac{\partial G}{\partial \Phi_j^{(i)}} - \frac{\partial}{\partial x_k}\frac{\partial G}{\partial \Phi_{j,k}^{(i)}} = 0\,, \tag{S.10a}$$

which have the following form

$$2b_{jl}^{(ik)}\Phi_l^{(k)} + 4b_{jlqr}^{(ikmn)}\Phi_l^{(k)}\Phi_q^{(m)}\Phi_r^{(n)} + 2\eta_{jlqr}^{(ikmn)}\Phi_l^{(k)}P_q^{(m)}P_r^{(n)} - 2\tilde{R}_{mnjl}^{(ik)}u_{mn}\Phi_l^{(k)} - \lambda^{(im)}\varepsilon_{kjl}\frac{\partial P_k^{(m)}}{\partial x_l} +$$

$$\xi_{ljmq}^{(kin)}P_l^{(k)}\frac{\partial \Phi_m^{(n)}}{\partial x_q} - \xi_{lmjq}^{(kni)}\frac{\partial\left[P_l^{(k)}\Phi_m^{(n)}\right]}{\partial x_q} - v_{jkqr}^{(il)}\frac{\partial^2\Phi_q^{(l)}}{\partial x_k \partial x_r} = 0. \tag{S.10b}$$

It is obvious that the conditions of the sublattices equivalence should be fulfilled, i.e., the appropriate renumbering of the sublattices should have no effect on the free energy. Below we restrict ourselves to the most common case of symmetry with respect to the cyclic transmutations of the sublattices only, namely:

$$a_{kl}^{(11)} \equiv a_{kl}^{(22)} \equiv a_{kl}^{(33)} \equiv a_{kl}^{(44)},\, a_{kl}^{(12)} \equiv a_{kl}^{(23)} \equiv a_{kl}^{(34)} \equiv a_{kl}^{(41)},\, a_{klqr}^{(1122)} \equiv a_{klqr}^{(2233)} \equiv a_{klqr}^{(3344)} \;... \tag{S.11}$$

Considering the cubic symmetry of the parent phase and the symmetry relations like Eq.(S.11), one obtains from Eq.(S.1) the following expressions for free energy contributions:

$$G_{AFD} = b^{(11)}\left(\boldsymbol{\Phi}^{(i)},\boldsymbol{\Phi}^{(i)}\right) + 2b^{(12)}\left(\boldsymbol{\Phi}^{(i)},\boldsymbol{\Phi}^{(i+1)}\right) + 2b^{(13)}\left(\boldsymbol{\Phi}^{(i)},\boldsymbol{\Phi}^{(i+2)}\right) + 3b_{1122}^{(1111)}\left|\boldsymbol{\Phi}^{(i)}\right|^2\left|\boldsymbol{\Phi}^{(i)}\right|^2 + \left\{b_{1111}^{(1111)} - 3b_{1122}^{(1111)}\right\}\sum_{i=1}^4\left[\left(\Phi_1^{(i)}\right)^4 + \left(\Phi_2^{(i)}\right)^4 + \left(\Phi_3^{(i)}\right)^4\right] + b_{1122}^{(1122)}\left|\boldsymbol{\Phi}^{(i)}\right|^2\left|\boldsymbol{\Phi}^{(i+1)}\right|^2 + 2b_{1212}^{(1122)}\left(\boldsymbol{\Phi}^{(i)},\boldsymbol{\Phi}^{(i+1)}\right)^2 + \left\{b_{1111}^{(1122)} - b_{1122}^{(1122)} - 2b_{1212}^{(1122)}\right\}\left[\left(\Phi_1^{(i)}\Phi_1^{(i+1)}\right)^2 + \left(\Phi_2^{(i)}\Phi_2^{(i+1)}\right)^2 + \left(\Phi_3^{(i)}\Phi_3^{(i+1)}\right)^2\right] + b_{1122}^{(1133)}\left|\boldsymbol{\Phi}^{(i)}\right|^2\left|\boldsymbol{\Phi}^{(i+1)}\right|^2 + 2b_{1212}^{(1133)}\left(\boldsymbol{\Phi}^{(i)},\boldsymbol{\Phi}^{(i+2)}\right)^2 + \left\{b_{1111}^{(1133)} - b_{1122}^{(1133)} - 2b_{1212}^{(1133)}\right\}\left[\left(\Phi_1^{(i)}\Phi_1^{(i+2)}\right)^2 + \left(\Phi_2^{(i)}\Phi_2^{(i+2)}\right)^2 + \left(\Phi_3^{(i)}\Phi_3^{(i+2)}\right)^2\right] \tag{S.12a}$$

$$G_{EP} = a^{(11)}\left(\mathbf{P}^{(i)},\mathbf{P}^{(i)}\right) + a^{(12)}\left(\mathbf{P}^{(i)},\mathbf{P}^{(i+1)}\right) + a^{(13)}\left(\mathbf{P}^{(i)},\mathbf{P}^{(i+2)}\right) + 3a_{1122}^{(1111)}\left|\mathbf{P}^{(i)}\right|^2\left|\mathbf{P}^{(i)}\right|^2 + \left\{a_{1111}^{(1111)} - 3a_{1122}^{(1111)}\right\}\sum_{i=1}^4\left[\left(P_1^{(i)}\right)^4 + \left(P_2^{(i)}\right)^4 + \left(P_3^{(i)}\right)^4\right] + a_{1122}^{(1122)}\left|\mathbf{P}^{(i)}\right|^2\left|\mathbf{P}^{(i+1)}\right|^2 + 2a_{1212}^{(1122)}\left(\mathbf{P}^{(i)},\mathbf{P}^{(i+1)}\right)^2 + \left\{a_{1111}^{(1122)} - a_{1122}^{(1122)} - 2a_{1212}^{(1122)}\right\}\left[\left(P_1^{(i)}P_1^{(i+1)}\right)^2 + \left(P_2^{(i)}P_2^{(i+1)}\right)^2 + \left(P_3^{(i)}P_3^{(i+1)}\right)^2\right] + a_{1122}^{(1133)}\left|\mathbf{P}^{(i)}\right|^2\left|\mathbf{P}^{(i+1)}\right|^2 + 2a_{1212}^{(1133)}\left(\mathbf{P}^{(i)},\mathbf{P}^{(i+2)}\right)^2 + \left\{a_{1111}^{(1133)} - a_{1122}^{(1133)} - 2a_{1212}^{(1133)}\right\}\left[\left(P_1^{(i)}P_1^{(i+2)}\right)^2 + \left(P_2^{(i)}P_2^{(i+2)}\right)^2 + \left(P_3^{(i)}P_3^{(i+2)}\right)^2\right] \tag{S.12b}$$

$$G_{HLS} = \eta_{1122}^{(1111)}\left|\boldsymbol{\Phi}^{(i)}\right|^2\left|\boldsymbol{P}^{(i)}\right|^2 + 2\eta_{1212}^{(1111)}\left(\boldsymbol{\Phi}^{(i)},\boldsymbol{P}^{(i)}\right)^2 + \left\{\eta_{1111}^{(1111)} - \eta_{1122}^{(1111)} - 2\eta_{1212}^{(1111)}\right\}\left[\left(\Phi_1^{(i)}P_1^{(i)}\right)^2 + \left(\Phi_2^{(i)}P_2^{(i)}\right)^2 + \left(\Phi_3^{(i)}P_3^{(i)}\right)^2\right] + \eta_{1122}^{(1122)}\left|\boldsymbol{\Phi}^{(i)}\right|^2\left|\mathbf{P}^{(i+1)}\right|^2 + 2\eta_{1212}^{(1122)}\left(\boldsymbol{\Phi}^{(i)},\boldsymbol{P}^{(i+1)}\right)^2 + \left\{\eta_{1111}^{(1122)} - \eta_{1122}^{(1122)} - 2\eta_{1212}^{(1122)}\right\}\left[\left(\Phi_1^{(i)}P_1^{(i+1)}\right)^2 + \left(\Phi_2^{(i)}P_2^{(i+1)}\right)^2 +\right.$$

$$\left(\Phi_3^{(i)}P_3^{(i+1)}\right)^2\Big] + \eta_{1122}^{(1133)}\left|\boldsymbol{\Phi}^{(i)}\right|^2\left|\mathbf{P}^{(i+2)}\right|^2 + 2\eta_{1212}^{(1133)}\left(\boldsymbol{\Phi}^{(i)},\boldsymbol{P}^{(i+2)}\right)^2 + \left\{\eta_{1111}^{(1133)} - \eta_{1122}^{(1133)} - 2\eta_{1212}^{(1133)}\right\}\left[\left(\Phi_1^{(i)}P_1^{(i+2)}\right)^2 + \left(\Phi_2^{(i)}P_2^{(i+2)}\right)^2 + \left(\Phi_3^{(i)}P_3^{(i+2)}\right)^2\right] + \eta_1^{(1212)}\left(\boldsymbol{\Phi}^{(\mathrm{i})},\boldsymbol{\Phi}^{(\mathrm{i+1})}\right)\left(\mathbf{P}^{(\mathrm{i})},\mathbf{P}^{(\mathrm{i+1})}\right) + \eta_2^{(1212)}\left(\boldsymbol{\Phi}^{(\mathrm{i})},\boldsymbol{P}^{(\mathrm{i+1})}\right)\left(\mathbf{P}^{(\mathrm{i})},\boldsymbol{\Phi}^{(\mathrm{i+1})}\right) + \eta_3^{(1212)}\left(\boldsymbol{\Phi}^{(\mathrm{i})},\boldsymbol{P}^{(\mathrm{i})}\right)\left(\mathbf{P}^{(\mathrm{i+1})},\boldsymbol{\Phi}^{(\mathrm{i+1})}\right) + \eta_{11}^{(1313)}\left(\boldsymbol{\Phi}^{(\mathrm{i})},\boldsymbol{\Phi}^{(\mathrm{i+2})}\right)\left(\mathbf{P}^{(\mathrm{i})},\mathbf{P}^{(\mathrm{i+2})}\right) + \eta_2^{(1313)}\left(\boldsymbol{\Phi}^{(\mathrm{i})},\mathbf{P}^{(\mathrm{i+2})}\right)\left(\mathbf{P}^{(\mathrm{i})},\boldsymbol{\Phi}^{(\mathrm{i+2})}\right) + \eta_3^{(1313)}\left(\boldsymbol{\Phi}^{(\mathrm{i})},\mathbf{P}^{(\mathrm{i})}\right)\left(\mathbf{P}^{(\mathrm{i+2})},\boldsymbol{\Phi}^{(\mathrm{i+2})}\right) \quad \text{(S.12c)}$$

$$G_{ST} = -\sum_{k,l,i,j} u_{mn}\left(Q_{mnkl}^{(11)}P_k^{(i)}P_l^{(i)} + Q_{mnkl}^{(12)}P_k^{(i)}P_l^{(i+1)} + Q_{mnkl}^{(13)}P_k^{(i)}P_l^{(i+2)} + R_{mnkl}^{(11)}\Phi_k^{(i)}\Phi_l^{(i)} + R_{mnkl}^{(12)}\Phi_k^{(i)}\Phi_l^{(i+1)} + R_{mnkl}^{(13)}\Phi_k^{(i)}\Phi_l^{(i+2)}\right) \quad \text{(S.12d)}$$

Equations of state for polarization and AFD order parameters of sublattices could be found from Euler-Lagrange equation, similarly to Eq.(S.9a) and (S.10a):

$$2a^{(11)}P_j^{(i)} + a^{(12)}\left[P_j^{(i-1)} + P_j^{(i+1)}\right] + a^{(13)}P_j^{(i+2)} + \Big[6a_{1122}^{(1111)}\left|\mathbf{P}^{(i)}\right|^2 + 4\left\{a_{1111}^{(1111)} - 3a_{1122}^{(1111)}\right\}\left(P_j^{(i)}\right)^2 + 2a_{1122}^{(1122)}\left[\left|\mathbf{P}^{(i-1)}\right|^2 + \left|\mathbf{P}^{(i+1)}\right|^2\right] + 2\left\{a_{1111}^{(1122)} - a_{1122}^{(1122)} - 2a_{1212}^{(1122)}\right\}\left[\left(P_j^{(i+1)}\right)^2 + \left(P_j^{(i-1)}\right)^2\right] + \ldots + 2\eta_{1122}^{(1111)}\left|\boldsymbol{\Phi}^{(i)}\right|^2 + 4\eta_{1212}^{(1111)}\left(\boldsymbol{\Phi}^{(i)},\boldsymbol{P}^{(i)}\right)\Big]P_j^{(i)} + 2a_{1212}^{(1122)}\left[\left(\mathbf{P}^{(i)},\mathbf{P}^{(i+1)}\right)P_j^{(i+1)} + \left(\mathbf{P}^{(i-1)},\mathbf{P}^{(i)}\right)P_j^{(i-1)}\right] - f_{mkjl}^{(i)}\frac{\partial u_{mk}}{\partial x_l} + \lambda^{(im)}\varepsilon_{jkl}\frac{\partial \Phi_k^{(m)}}{\partial x_l} + \xi_{jmnq}^{(ilk)}\Phi_m^{(l)}\frac{\partial \Phi_n^{(k)}}{\partial x_q} - g_{jlqr}^{(ik)}\frac{\partial^2 P_q^{(k)}}{\partial x_l \partial x_r} = 0 \quad \text{(S.13a)}$$

$$2b^{(11)}\Phi_j^{(i)} + b^{(12)}\left(\Phi_j^{(i-1)} + \Phi_j^{(i+1)}\right) + b^{(13)}\Phi_j^{(i+2)} + \Big[4b^{(1111)}\left|\boldsymbol{\Phi}^{(i)}\right|^2 + 2b^{(1122)}\left|\boldsymbol{\Phi}^{(i+1)}\right|^2 + \ldots + 2\eta^{(1111)}\left|\mathbf{P}^{(i)}\right|^2 + 2\eta^{(1122)}\left|\mathbf{P}^{(i+2)}\right|^2\Big]\Phi_j^{(i)} - \lambda^{(im)}\varepsilon_{kjl}\frac{\partial P_k^{(m)}}{\partial x_l} + \xi_{ljmq}^{(kin)}P_l^{(k)}\frac{\partial \Phi_m^{(n)}}{\partial x_q} - \xi_{lmjq}^{(kni)}\frac{\partial\left[P_l^{(k)}\Phi_m^{(n)}\right]}{\partial x_q} - v_{jkqr}^{(il)}\frac{\partial^2 \Phi_q^{(l)}}{\partial x_k \partial x_r} = 0 \quad \text{(S.13b)}$$

Below the rotated frame with the new coordinates are introduced:

$$\tilde{x}_1 = \frac{1}{\sqrt{2}}(x_1 + x_2),\ \tilde{x}_2 = \frac{1}{\sqrt{2}}(x_2 - x_1) \text{ and } \tilde{x}_3 = x_3 \quad \text{(S.14)}$$

In the rotated coordinate frame, the coupling tensor components have the following form

$$\tilde{\xi}_{1111} = \frac{1}{4}(\xi_{1111} + \xi_{1122} + \xi_{1212} + \xi_{1221} + \xi_{2112} + \xi_{2121} + \xi_{2211} + \xi_{2222}), \quad \text{(S.15a)}$$

$$\tilde{\xi}_{1221} = \frac{1}{4}(\xi_{1111} - \xi_{1122} - \xi_{1212} + \xi_{1221} + \xi_{2112} - \xi_{2121} - \xi_{2211} + \xi_{2222}), \quad \text{(S.15b)}$$

$$\tilde{\xi}_{2121} = \frac{1}{4}(\xi_{1111} - \xi_{1122} + \xi_{1212} - \xi_{1221} - \xi_{2112} + \xi_{2121} - \xi_{2211} + \xi_{2222}), \quad \text{(S.15c)}$$

$$\tilde{\xi}_{2211} = \frac{1}{4}(\xi_{1111} + \xi_{1122} - \xi_{1212} - \xi_{1221} - \xi_{2112} - \xi_{2121} + \xi_{2211} + \xi_{222}). \quad \text{(S.15d)}$$

Considering the cubic symmetry of the parent phase, i.e., $\xi_{1111} \equiv \xi_{2222}$, $\xi_{1122} \equiv \xi_{2211}$, $\xi_{1212} \equiv \xi_{2121}$, $\xi_{1221} \equiv \xi_{2112}$, one could rewrite the relations (S.15) as follows

$$\tilde{\xi}_{1111} = \frac{1}{2}(\xi_{1111} + \xi_{1122} + \xi_{1212} + \xi_{1221}), \quad \tilde{\xi}_{1221} = \frac{1}{2}(\xi_{1111} - \xi_{1122} - \xi_{1212} + \xi_{1221}), \quad \text{(S.16a)}$$

$$\tilde{\xi}_{2121} = \frac{1}{2}(\xi_{1111} - \xi_{1122} + \xi_{1212} - \xi_{1221}), \quad \tilde{\xi}_{2211} = \frac{1}{2}(\xi_{1111} + \xi_{1122} - \xi_{1212} - \xi_{1221}). \quad \text{(S.16b)}$$

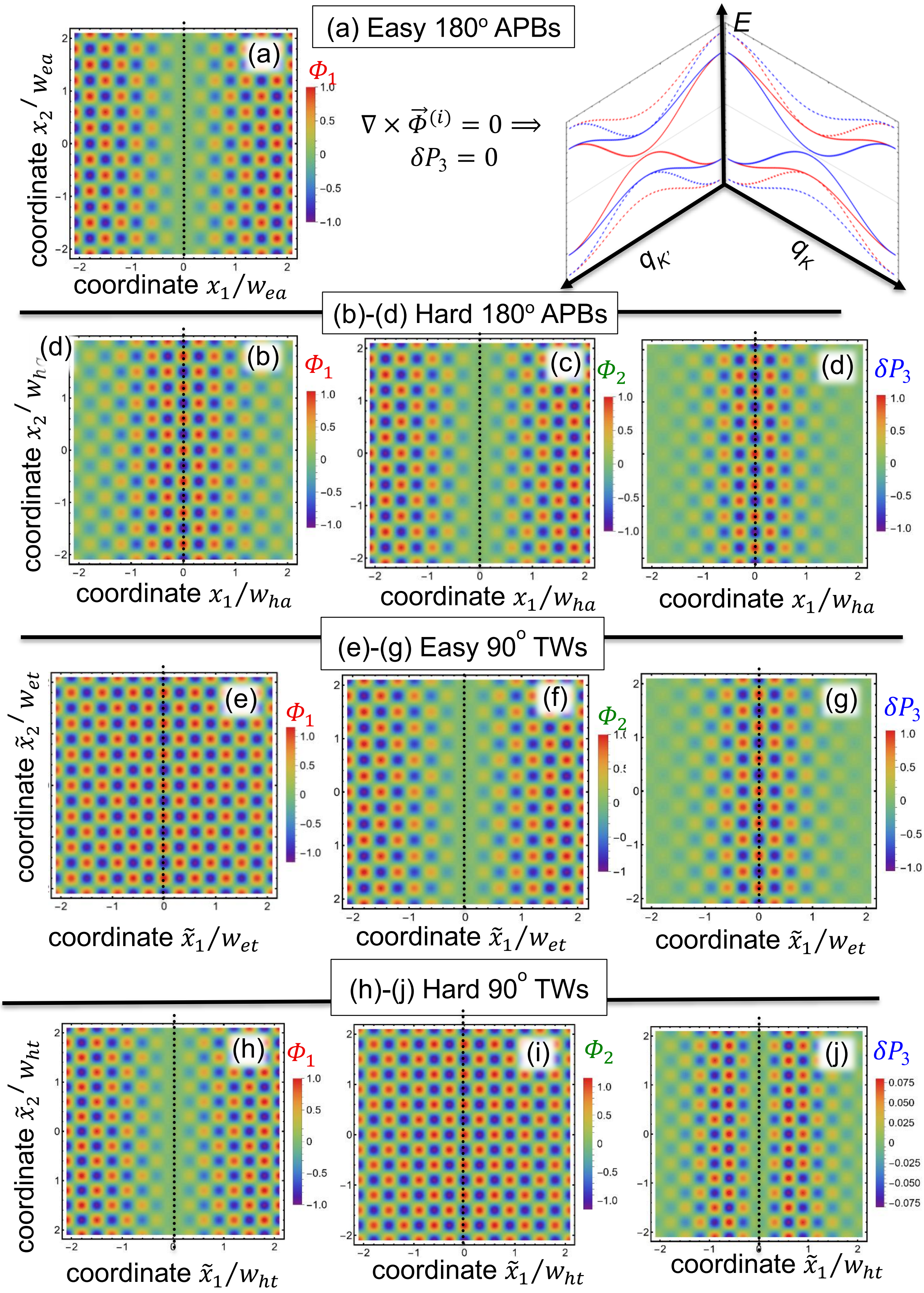

**FIGURE S1.** Color maps of the tilt components $\widetilde{\Phi}_1^{(i)}$ and $\widetilde{\Phi}_2^{(i)}$, and polarization component $\delta\widetilde{P}_3^{(i)}$, which are checkerboard-type sign-alternating functions of $\tilde{x}_1$ and $\tilde{x}_2$.

**Supplement S2**. **Dispersion laws for the free carriers near the structural domain walls**

Within the Bardeen-Shockley-Pikus-Bir deformation potential approach, the strain-induced local changes of the band structure are determined by the Bir and Pikus [34] perturbation Hamiltonian,

$$\hat{H}_{PB}\left(u_{ij}^{f}\right) = \hat{D}_{ij}^{PB} u_{ij}^{f} \approx -a\, Tr\left[u_{ij}^{f}\right]\hat{I} - 3b \sum_i \left(\hat{J}_i^2 - \frac{1}{3}\hat{J}^2\right) u_{ii}^{f} - \frac{2d}{\sqrt{3}} \sum_{i \neq j} \left[\hat{J}_i, \hat{J}_j\right]_+ u_{ij}^{f}, \quad \text{(S.17a)}$$

where $\hat{D}_{ij}^{PB}$ is deformation potential operator and $u_{ij}^{f}(\vec{x})$ is the strain variation, which is assumed to be a slowly varying function of spatial coordinates compared to the lattice constant (e.g., the envelope).

In the simplest case considered analytically by Bir and Pikus, the valence band is triply degenerate in orbital angular momentum (p-like states with $l = 1$) in the center of the Brillouin zone ($\vec{k} = 0$). Considering the spin-orbit interaction, it splits into a heavy hole HH-band, a light hole LH-band with a total momentum $J = 3/2$ and a split spin-orbital (SO) band with $J = 1/2$. The approximate equality in Eq. (S.17a) is valid in the four-dimensional space of sub-band states with the total momentum $J = \frac{3}{2}$. Here $\hat{I}$ is the 4x4 unit matrix, $a$, $b$ and $d$ are the fundamental constants of deformation potential for valence (or conductive) band, $\hat{J}_i$ are the operators of angular momentum, $\hat{J}^2 = \sum_i \hat{J}_i^2 = J(J+1) = \frac{15}{4}\hat{I}$ and $\left[\hat{J}_i, \hat{J}_j\right]_+ = \frac{1}{2}\left(\hat{J}_i\hat{J}_j + \hat{J}_j\hat{J}_i\right)$ is the anticommutator of the momentum operator. Using the basis of the momentum operator $\hat{J}$ eigen functions, namely the functions $|3/2, 3/2\rangle$, $|3/2, 1/2\rangle$, $|3/2, -1/2\rangle$ and $|3/2, -3/2\rangle$, the approximate equality in Eq. (S.17a) can be written in the matrix form:

$$\hat{H}_{PB} \approx \begin{pmatrix} S_V + T_V & U_V & W_V & 0 \\ U_V^* & S_V - T_V & 0 & W_V \\ W_V^* & 0 & S_V - T_V & -U_V \\ 0 & W_V^* & -U_V^* & S_V + T_V \end{pmatrix}, \quad \text{(S.17b)}$$

The parameters

$$S_V = -a\left(u_{11}^{f} + u_{22}^{f} + u_{33}^{f}\right), \quad T_V = -b\left[u_{33}^{f} - \frac{1}{2}\left(u_{11}^{f} + u_{22}^{f}\right)\right], \quad \text{(S.17c)}$$

$$U_V = \frac{\sqrt{3}}{2} b\left(u_{11}^{f} - u_{22}^{f}\right) - i d u_{12}^{f}, \qquad W_V = -d\left(u_{13}^{f} - i u_{23}^{f}\right). \quad \text{(S.17d)}$$

The dispersion laws of the holes in the the valence band, $E_V(\vec{k})$, and electrons in the conduction band, $E_C(\vec{k})$, can be estimated from the effective mass approximation and perturbation approach. Using the approach, we consider the $\hat{H}_{PB}(\vec{x})$ as a perturbation to the Kohn-Luttinger Hamiltonian. The Kohn-Luttinger Hamiltonian $\hat{H}_{KL}(\vec{x})$ has the form

$$\hat{H}_{KL}(\vec{x}) = -\frac{\hbar^2}{2m_0}\begin{pmatrix} S_L + T_L & -U_L & W_L & 0 \\ -U_L^\dagger & S_L - T_L & 0 & W_L \\ W_L^\dagger & 0 & S_L - T_L & U_L \\ 0 & W_L^\dagger & U_L^\dagger & S_L + T_L \end{pmatrix}, \qquad \text{(S.18a)}$$

Here the operators are introduced:

$$S_L = \gamma_1\left(\frac{\partial^2}{\partial x_1^2} + \frac{\partial^2}{\partial x_2^2} + \frac{\partial^2}{\partial x_3^2}\right), \qquad \text{(S.18b)}$$

$$T_L = \gamma_2\left(\frac{\partial^2}{\partial x_1^2} + \frac{\partial^2}{\partial x_2^2} - 2\frac{\partial^2}{\partial x_3^2}\right), \qquad \text{(S.18c)}$$

$$U_L = -\sqrt{3}\left[\gamma_2\left(\frac{\partial^2}{\partial x_1^2} - \frac{\partial^2}{\partial x_2^2}\right) - 2i\gamma_3\frac{\partial^2}{\partial x_1 \partial x_2}\right], \qquad \text{(S.18d)}$$

$$W_L = 2\sqrt{3}\gamma_3\left(\frac{\partial^2}{\partial x_1 \partial x_3} - i\frac{\partial^2}{\partial x_2 \partial x_3}\right). \qquad \text{(S.18e)}$$

Here $m_0$ is the mass of free electron and $\gamma_i$ are Luttinger parameters, which determine the effective mass tensor. Eigen functions of the $\hat{H}_{KL}(\vec{x})$ are 4-component spinors,

$$\psi_k(\vec{x}) = -\frac{\exp(i\vec{k}\vec{r})}{\sqrt{V}}\hat{s}, \qquad \text{(S.18f)}$$

where $V$ is the actual volume, $\hat{s}$ is the 4-component vector-column spinor, which elements are the combinations of Bloch amplitudes corresponding to the states with the projection of the full moment $m_j = \pm\frac{3}{2}, \pm\frac{1}{2}$. The Bloch amplitudes are periodic functions, which period in the corresponding direction is the lattice parameter. In the "parabolic" approximation the $\hat{H}_{KL}$ asquires the form:

$$\hat{H}_{KL}(\vec{k}, \hat{J}_i) = -\frac{\hbar^2}{2m_0}\left[\gamma_1 k^2 + \gamma_2\left(\sum_i \hat{J}_i^2 k_i^2 - \frac{1}{2}\hat{J}^2 k^2\right) + 2\gamma_3 \sum_{i\neq j}\left(\hat{J}_i \hat{J}_j k_i k_j\right)\right]. \qquad \text{(S.19)}$$

where $k^2 = k_1^2 + k_2^2 + k_3^2$ and $\hat{J}^2 = \sum_i \hat{J}_i^2 = J(J+1) = \frac{15}{4}\hat{I}$.

Next, the full Hamiltonian, $\hat{H}_{KL}(\vec{x}) + \hat{H}_{PB}(u_{ij}^f)$, in k-space is given by the matrix operator

$$\hat{H}_L(\vec{k}) + \hat{H}_{PB}(u_{ij}^k) \cong \begin{pmatrix} S_{Vk} + T_{Vk} & U_{Vk} & W_{Vk} & 0 \\ U_{Vk}^* & S_{Vk} - T_{Vk} & 0 & W_{Vk} \\ W_{Vk}^* & 0 & S_{Vk} - T_{Vk} & -U_{Vk} \\ 0 & W_{Vk}^* & -U_{Vk}^* & S_{Vk} + T_{Vk} \end{pmatrix}. \qquad \text{(S.20a)}$$

Here the k-dependent parameters are introduced:

$$S_{Vk} = -\frac{\hbar^2}{2m_0}\gamma_1(k_1^2 + k_2^2 + k_3^2) - a\left(u_{11}^k + u_{22}^k + u_{33}^k\right), \qquad \text{(S.20b)}$$

$$T_{Vk} = -\frac{\hbar^2}{2m_0}\gamma_2(k_1^2 + k_2^2 - 2k_3^2) - b\left[u_{33}^k - \frac{1}{2}\left(u_{11}^k + u_{22}^k\right)\right], \qquad \text{(S.20c)}$$

$$U_{Vk} = -\frac{\sqrt{3}\hbar^2}{2m_0}[\gamma_2(k_1^2 - k_2^2) - 2i\gamma_3 k_1 k_2] + \frac{\sqrt{3}}{2}b\left(u_{11}^k - u_{22}^k\right) - idu_{12}^k, \qquad \text{(S.20d)}$$

$$W_{Vk} = -\sqrt{3}\frac{\hbar^2\gamma_3}{m_0}(k_1 - ik_2)k_3 - d\left(u_{13}^k - iu_{23}^k\right). \qquad \text{(S.20e)}$$

The dispersion laws for light and heavy holes can be derived from the condition of zero determinant of the matrix operator (S.20a) that has the form:

$$E_V(\vec{k}) = E_{V0} + S_{Vk} \pm \sqrt{T_{Vk}^2 + |U_{Vk}^2| + |W_{Vk}^2|}. \qquad \text{(S.21a)}$$

A similar condition can be derived for electrons in the conductive band:

$$E_C(\vec{k}) = E_{C0} + S_{Ck} \pm \sqrt{T_{Ck}^2 + \left|U_{Ck}^2\right| + \left|W_{Ck}^2\right|}. \qquad \text{(S.21b)}$$

The values $E_{V0}$ and $E_{C0}$ are the bands energetic positions far from the TWs and APBs. Explicit form of Eqs. (S.21) depends on the explicit form of the strains $u_{ij}^k$. The derivation of Eq. (S.21) is rigorous for coordinate-independent strains, namely for $u_{ij}^k \equiv u_{ij}^f = const$. Otherwise, the strain-related parameters $u_{ij}^k$ should be calculated using the proper averaging and perturbation approach. In the first order of perturbation theory:

$$u_{ij}^k = \langle \psi_k | \hat{u}_{ij}^f | \psi_k \rangle, \qquad \text{(S.21c)}$$

where the operator $\hat{u}_{ij}^f$ correspond to the acoustic phonon modes influenced by the flexo-AFD coupling.

According to Eqs.(S.20b-e) the dilatational strains $u_{ii}^k$ shift the bands as a whole and do not introduce any asymmetry with respect to the sign of $\vec{k}$. The shear strains $u_{ij}^k$ $(i \neq j)$ can change the band structure in an asymmetric way.

In what follows we are interested to estimate the influence of the last terms in Eq. (9a), $\delta u_{ij}^f = \left[Q_{ijkl}^{(mn)} P_l^{(n)} - F_{ijkl}^{(m)} \frac{\partial}{\partial x_l}\right] P_k^{(m)}$ on the local band structure, since these terms contains linear and quadratic contributions proportional to $\delta P_{03}^{(m)} \sim \chi \Phi_s$ and $\left(\delta P_{03}^{(m)}\right)^2 \sim (\chi \Phi_s)^2$, respectively. The contributions are induced by the bilinear flexo-AFD coupling.

For a cubic symmetry of the parent phase, nonzero components of the electrostriction and flexoelectric tensor are $Q_{1111}^{(m)} = Q_{2222}^{(m)} = Q_{3333}^{(m)}$, $Q_{1212}^{(m)} = Q_{2323}^{(m)} = Q_{1313}^{(m)}$ and $Q_{1122}^{(m)} = Q_{1133}^{(m)} = Q_{2233}^{(m)}$; $F_{1111}^{(m)} = F_{2222}^{(m)} = F_{3333}^{(m)}$, $F_{1212}^{(m)} = F_{2323}^{(m)} = F_{1313}^{(m)}$ and $F_{1122}^{(m)} = F_{1133}^{(m)} = F_{2233}^{(m)}$. The nonzero components of the strain variations are $\delta u_{33}^f = \tilde{Q}_{3333}^{(mn)} \left(\delta \tilde{P}_{03}^{(m)}\right)^2$, $\delta u_{11}^f = \delta u_{22}^f = \tilde{Q}_{1133}^{(mn)} \left(\delta \tilde{P}_{03}^{(m)}\right)^2$, $\delta u_{12}^f = -\tilde{F}_{1212}^{(m)} \frac{\partial \delta \tilde{P}_{02}^{(m)}}{\partial \tilde{x}_1}$, $\delta u_{23}^f = -\tilde{F}_{3232}^{(m)} \frac{\partial \delta \tilde{P}_{03}^{(m)}}{\partial \tilde{x}_2}$, and $\delta u_{31}^f = -\tilde{F}_{3131}^{(m)} \frac{\partial \delta \tilde{P}_{03}^{(m)}}{\partial \tilde{x}_1}$ for TWs; $\delta u_{33}^f = Q_{3333}^{(mn)} \left(\delta P_{03}^{(m)}\right)^2$, $\delta u_{11}^f = \delta u_{22}^f = Q_{1133}^{(mn)} \left(\delta P_{03}^{(m)}\right)^2$ and $\delta u_{31}^f = -F_{3131}^{(m)} \frac{\partial \delta P_{03}^{(m)}}{\partial x_1}$ for APBs. Only the slowly varying envelope functions $P_{03}$ (or their even powers) matter in these expressions. The oscillating functions proportional to $\exp(i k_0 x_1)$ or $\exp(i \tilde{k}_0 \tilde{x}_1)$, where $\tilde{k}_0 \cong \frac{2\pi}{\tilde{a}_0}$ or $k_0 \cong \frac{2\pi}{a_0}$ are inverse lattice vectors in the corresponding directions of the Brillouin zone, should be included to the Bloch amplitudes. Only the smooth envelope functions can change the band structure.

As a concrete example, let us consider a "toy model" case $\gamma_2 = 0$ and $b = 0$. Equations (S.20) and (S.21) acquire the simple form

$$\hat{H}_L(\vec{k}) + \hat{H}_{PB}(u_{ij}^k) \cong \begin{pmatrix} S_{Vk} & U_{Vk} & W_{Vk} & 0 \\ U_{Vk}^* & S_{Vk} & 0 & W_{Vk} \\ W_{Vk}^* & 0 & S_{Vk} & -U_{Vk} \\ 0 & W_{Vk}^* & -U_{Vk}^* & S_{Vk} \end{pmatrix}. \tag{S.22a}$$

$$S_{Vk} = -\frac{\hbar^2}{2m_0}\gamma_1(k_1^2 + k_2^2 + k_3^2) - a(u_{11}^k + u_{22}^k + u_{33}^k), \tag{S.22b}$$

$$U_{Vk} = \frac{\sqrt{3}\hbar^2}{m_0} i\gamma_3 k_1 k_2 - idu_{12}^k, \tag{S.22c}$$

$$W_{Vk} = -\sqrt{3}\frac{\hbar^2\gamma_3}{m_0}(k_1 - ik_2)k_3 - d(u_{13}^k - iu_{23}^k). \tag{S.22d}$$

Therefore, for the case $u_{ij}^k \neq 0$, the dispersion law Eq. (S.21) acquires the form

$$E_V(\vec{k}) = E_{V0} - \frac{\hbar^2}{2m_0}\gamma_1\left[k_1^2 + k_2^2 + k_3^2 - \frac{2m_0 a}{\hbar^2}(u_{11}^k + u_{22}^k + u_{33}^k)\right] \pm$$

$$\frac{\sqrt{3}\hbar^2}{m_0}\gamma_3\sqrt{\left(k_1k_2 - \frac{m_0 d}{\sqrt{3}\hbar^2\gamma_3}u_{12}^k\right)^2 + \left(k_2k_3 - \frac{m_0 d}{\sqrt{3}\hbar^2\gamma_3}u_{23}^k\right)^2 + \left(k_1k_3 - \frac{m_0 d}{\sqrt{3}\hbar^2\gamma_3}u_{13}^k\right)^2}. \tag{S.23}$$

In the plane $k_2 = 0$, and for the case $u_{13}^k \neq 0$, Eq. (S.23) yields:

$$E_V(\vec{k}) = E_{V0} - \frac{\hbar^2}{2m_0}\gamma_1(k_1^2 + k_3^2) \pm \frac{\sqrt{3}\hbar^2}{m_0}\gamma_3\left|k_1k_3 - D_f u_{13}^k\right|, \tag{S.24a}$$

$$E_C(\vec{k}) = E_{C0} + \frac{\hbar^2}{2m_0}\gamma_1(k_1^2 + k_3^2) \mp \frac{\sqrt{3}\hbar^2}{m_0}\gamma_3\left|k_1k_3 - D_f u_{13}^k\right|, \tag{S.24b}$$

where the parameter $D_f = \frac{m_0 d}{\sqrt{3}\hbar^2\gamma_3}$ is introduced. Using the elastic dipole tensor of $SrTiO_3$ [35], the shear component $d$ can be estimated as $d \cong 3 - 30$ eV, and thus the parameter $D_f$ can be estimated as 9 – 90 nm$^{-2}$ for $\gamma_3 = 3$, $\hbar = 6.62\cdot10^{-34}$ J·s and $m_0 = 9.11\cdot10^{-31}$ kg. Thus, the product $D_f u_{13}^k \sim$ 0.01 - 0.9 nm$^{-2}$ for $u_{13}^k \cong 0.01$.

Let us postpone more rigorous calculations to further works, and analyze the model case of Tailor series expansion for the $u_{13}^k$:

$$u_{13}^k(k) \approx u_{13}^{(0)} + k_1 u_{13}^{(1)} + k_3 u_{13}^{(3)} + k_1k_3 u_{13}^{(13)} \sim (u_0 + Ak_1 + Bk_3 + Ck_1k_3)P_{03}, \tag{S.25}$$

where the wavenumber $k_1$ and $k_3$ originate from the near-wall acoustic phonon mode affected by the flexoelectric coupling. Equations (S.24) can be rewritten as

$$E_V(\vec{k}) = E_{V0} - \frac{\hbar^2}{2m_p}\left(k_1^2 + k_3^2 \pm \left|g_p k_1 k_3 - p_p - f_p k_1 - h_p k_3\right|\right), \tag{S.26a}$$

$$E_C(\vec{k}) = E_{C0} + \frac{\hbar^2}{2m_e}\left(k_1^2 + k_3^2 \pm \left|g_n k_1 k_3 - p_n - f_n k_1 - h_n k_3\right|\right), \tag{S.26b}$$

where the effective masses, $m_p = \frac{m_0}{\gamma_1^p}$ and $m_n = \frac{m_0}{\gamma_1^n}$, the strain-renormalized ratios of Luttinger parameters, $g_n = \sqrt{3}\frac{\gamma_3^n}{\gamma_1^n}(1 + C_n P_{03})$ and $g_p = \sqrt{3}\frac{\gamma_3^p}{\gamma_1^p}\left(1 + C_p P_{03}\right)$, are introduced. Other parameters are: $p_n = \frac{2m_0 d}{\hbar^2}\frac{\gamma_3^n}{\gamma_1^n}u_0 \sim \frac{2m_0 d}{\hbar^2}\frac{\gamma_3^n}{\gamma_1^n}P_{03}$, $f_n = \frac{2m_0 d}{\hbar^2}\frac{\gamma_3^n}{\gamma_1^n}A_n P_{03}$, $h_n = \frac{2m_0 d}{\hbar^2}\frac{\gamma_3^n}{\gamma_1^n}B_n P_{03}$, $p_p = \frac{2m_0 d}{\hbar^2}\frac{\gamma_3^p}{\gamma_1^p}u_0$, $f_p = \frac{2m_0 d}{\hbar^2}\frac{\gamma_3^p}{\gamma_1^p}A_p P_{03}$and $h_p = \frac{2m_0 d}{\hbar^2}\frac{\gamma_3^p}{\gamma_1^p}B_p P_{03}$. These parameters are proportional to the polarization $P_{03}$ and thus changes their sign with the sign of $P_{03}$.

A schematic picture of the local band bending calculated near structural walls in the AFD ferroelastics without and with prevailing bilinear flexo-AFD coupling is shown in **Figs. S2(a)** and **S2(b)**, respectively. It is seen from **Figs. S2(b)**, that the bilinear flexo-AFD coupling leads to the Rashba-Edelstein type splitting of the parabolic sub-bands with respect to the sign of $\delta P_{03}^{(m)} \sim \chi \Phi_s$ near the structural domain boundaries. Namely, the horizontal shift of red and blue parabolas is proportional to the parameter $f_p$ for holes in the valence band, or to the parameter $f_n$ for electrons in the conductive band. The parameters $g_n$ and $g_p$ determine the difference between the dispersion of light and easy carriers.

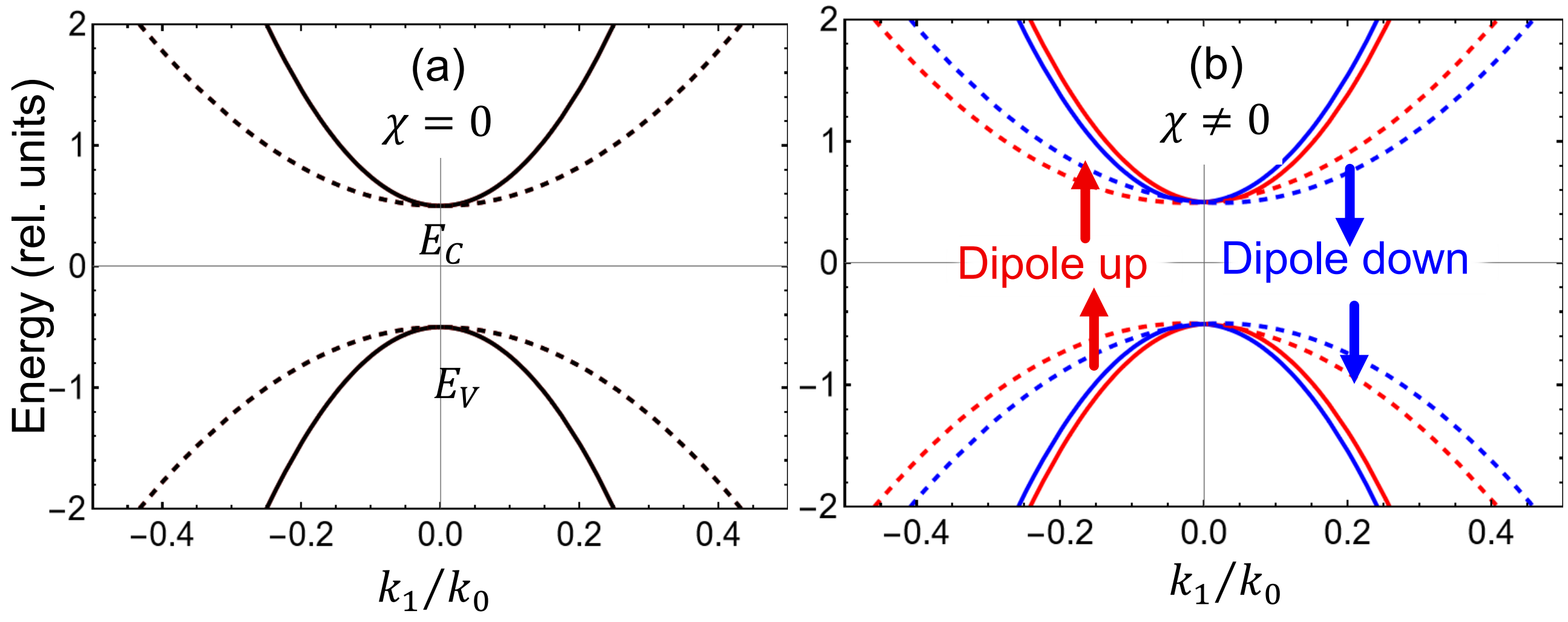

**FIGURE S2.** Schematic picture of the local band bending, $E_V(\vec{k})$ and $E_C(\vec{k})$ (in normalized units) near structural walls in the AFD ferroelastics without **(a)** and with prevailing **(b)** bilinear flexo-AFD coupling. Solid and dashed curves, which correspond to the dispersion curves of light and heavy carriers, respectively, are calculated from Eqs. (S.26) for parameters $C_n = C_p = 0$, $h_p = h_n = 0$, $p_n = p_p = 0$, $g_p = g_n = 0.5$, $f_n = f_p = 5$ (for red curves), $f_n = f_p = -5$ (for blue curves), $k_2 = 0$, and $k_3 = k_1$. The parameter $k_0$ is the inverse lattice vector in corresponding direction of the Brillouin space.